\documentclass[aps,prb,twocolumn,amsmath,amssymb,floatfix,superscriptaddress]{revtex4-1}
\usepackage{graphicx}
\usepackage{bm}
\usepackage{hyperref}
\usepackage{float}
\usepackage{xcolor}
\usepackage[utf8]{inputenc}

\usepackage{indentfirst}    
\begin{document}

\title{Field-free diode effects in one-dimensional superconductor: a complex  
interplay between Fulde-Ferrell pairing and 
altermagnetism  }

\author{SVS Sai Ruthvik}
\affiliation{Department of Physics, BITS Pilani-Hyderabad Campus, Telangana 500078, India}

\author{Tanay Nag}
\email{tanay.nag@hyderabad.bits-pilani.ac.in}
\affiliation{Department of Physics, BITS Pilani-Hyderabad Campus, Telangana 500078, India}

\begin{abstract}
We investigate the emergence of nonreciprocal dissipationless supercurrents in one-dimension manifested through the superconducting diode effect (SDE) and the Josephson diode effect (JDE) in the absence of any external magnetic field, where inversion symmetry (IS) and time-reversal symmetry (TRS) can be intrinsically broken by spin–orbit coupling (SOC), and altermagnetism (AM), respectively. We investigate  Ising and Rashba SOC separately in two models where two-component AM, parametrized by the crystallographic angle ($\alpha$), can lead to qualitatively similar indirect band-gap closing and non-reciprocal supercurrent in a Fulde-Ferrell (FF) superconductor. Interestingly, in the absence of SOCs, SDE persists, and the sign of efficiency can be altered solely by tuning the angle. Parallel spin components with complementary momentum functions, which break IS and TRS, can induce SDE with an efficiency of $\sim 80\%$ in the presence of FF pairing.  Continuing the analysis in the context of JDE, we explore the interplay between SOCs and AM with the  $p$-wave and Fulde-Ferrell superconductivity in three different setups.  The bulk bands contribute to the non-reciprocity in the case of $p$-wave superconductivity, whereas JDE is dominated by Andreev bound states in FF superconductivity, leading to efficiencies of $\sim 65\%$ and $30\%$ in the former and latter cases, respectively. Importantly, JDE continues to exist due to the finite momentum of the Cooper pair even without AM and SOC, unlike the $p$-wave superconductivity. The sign of JD efficiency can be tuned with angle in $p$-wave superconductivity, while the absence of such a sign reversal is a hallmark of FF superconductivity. Similar to SDE, parallel spin components, in conjunction with $p$-wave superconductivity, can lead to JDE that can also be mediated solely by FF pairing in the absence of SOC and AM. 

\end{abstract}

\maketitle

\section{Introduction}

Superconductivity is an extraordinary phase of matter in which many-body quantum effects cause electric current to flow without any resistance below a critical temperature. The celebrated Bardeen-Cooper-Schrieffer (BCS) pairing mechanism,  accompanied by zero center-of-mass momentum of the
Cooper pairs can explain the perfectly reciprocal, i.e., direction-independent behaviour of dissipationless supercurrent \cite{PhysRev.108.1175}. Recent experimental studies have shown that the non-reciprocity in the supercurrent can lead to a  diode effect namely, superconducting diode effect (SDE) \cite{ando2020observation,takeuchi2025superconducting,zhou2025superconducting,PhysRevB.107.054506,lyu2021superconducting,PhysRevLett.131.027001,li2025absence,PhysRevLett.134.236703,qi2025high}. There exist asymmetry between forward and reverse critical currents. 
In order to understand the above phenomena, the finite center-of-mass momentum of Cooper pairs is found to be a key ingredient while this type of unconventional pairing was theoretically predicted by Fulde and Ferrell (FF) \cite{PhysRev.135.A550,matsuda2007fulde,PhysRevB.72.184501,PhysRevB.76.014503,PhysRevLett.99.187002,PhysRevB.73.214514}. In the presence of Zeeman field and 
spin-orbit coupling (SOC) the Fermi-surface gets deformed allowing two opposite spins but with unequal momentum to pair up and form FF superconductor \cite{qu2013topological,zhang2013topological,yuan2021topological}.  Interestingly, there have been a significant amount of  theoretical endeavour to understand the exotic phenomena of SDE which is primarily mediated by the magnetic field, SOC and FF pairing \cite{Fu2022, Ando2020, Legg2022,PhysRevLett.128.037001,PhysRevB.106.205206,PhysRevB.109.094501,cv8s-tk4c,bhowmik2025optimizing,PhysRevB.111.L161402,ma2025superconducting,nadeem2023superconducting,PhysRevB.111.174514,angehrn2024relations,PhysRevB.106.214524,he2022phenomenological,karabassov2023superconducting,PhysRevB.110.024508}. It is important to note that SOC breaks inversion symmetry (IS) and magnetic field breaks time-reversal symmetry (TRS) leading to magnetochiral asymmetry (MCA) that is required to experience SDE \cite{nadeem2023superconducting,PhysRevLett.87.236602,PhysRevLett.130.136301,PhysRevB.99.245153}. Importantly, the interplay of magnetic field and SOC also leads to topological superconductivity, however, the deformation of the Fermi-surface and gap opening depends on the  mutual orientation between SOC and magnetic field \cite{PhysRevB.111.L121401,PhysRevB.107.035427,qu2013topological,zhang2013topological,yuan2021topological}. Apart from scientific viewpoint, technologically SDE is a promising phenomenon leading to various quantum devices such as SDE rectifier \cite{ingla2025efficient}.

\begin{figure}
    \centering
    \includegraphics[width=0.5\textwidth]{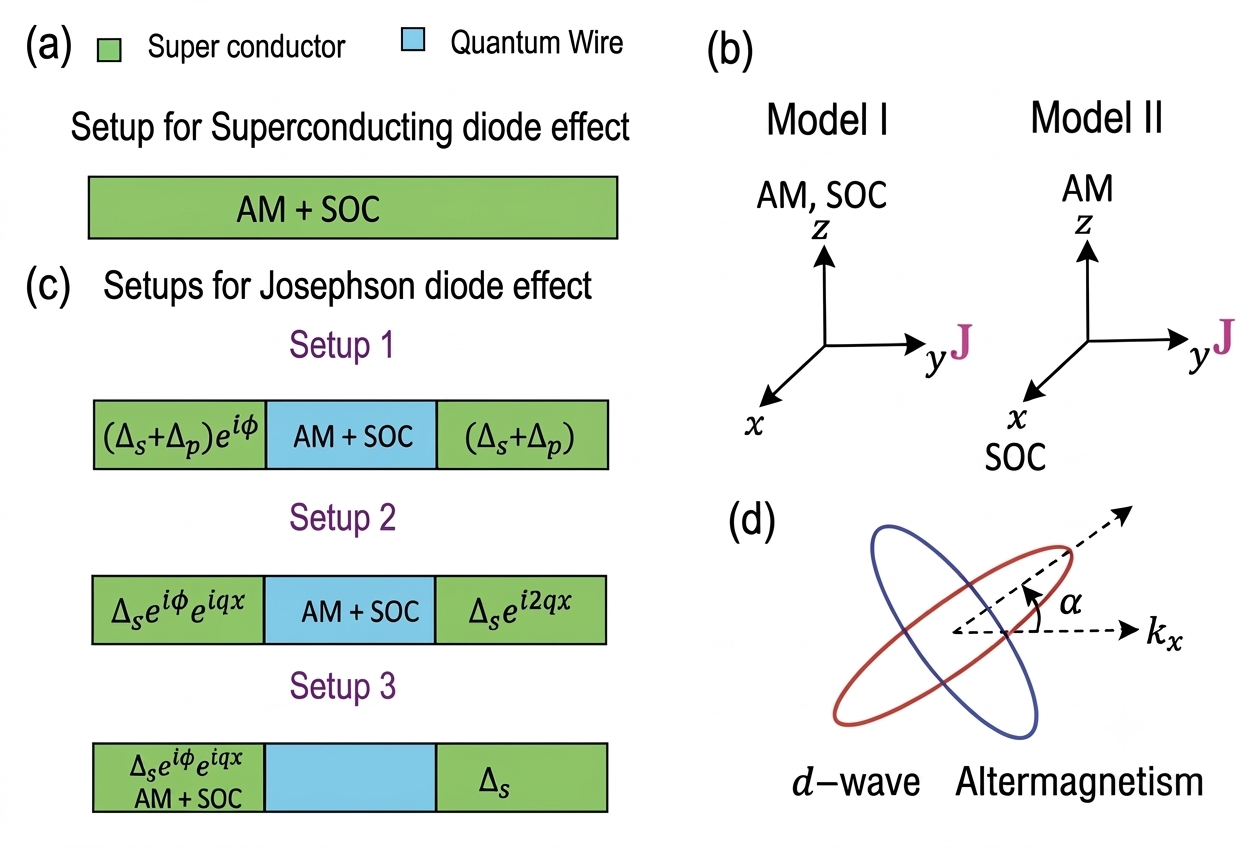}
    \caption{{We illustrate the setup for SDE in (a). The corresponding models, namely Model I, hosting parallel Ising SOC and AM,  and Model II, hosting transverse Rashba SOC and AM, are shown in (b). We demonstrate the JDE setups 1, 2,  and 3 in (c). The green and sky regions denote the superconducting and non-superconducting regions, respectively while  both of them host hopping terms.  In (d), we demonstrate the schematic of the Fermi surface of altermagnetism with a crystal angle, $\alpha$. }}
    \label{fig: Schematic}
\end{figure}

Going beyond the non-reciprocity of the  bulk supercurrent in SDE, there have been a great 
attention to study the nonreciprocal behavior of Josephson
current where two superconductors having different phases are interfaced together. The experimental emergence of Josephson diode effect (JDE) enriches the field of non-reciprocal transport in superconductors \cite{PhysRevB.110.104510,turini2022josephson,trahms2023diode,coraiola2024flux,gao2024josephson,wu2022field,wu2021realization,davydova2022universal,fracassi2024anomalous}. Interestingly, JDE along with finite Cooper pair momentum have been simultaneously observed experimentally  \cite{pal2022josephson}. Theoretical studies indicate towards the fact that the IS and TRS breaking are very important to obtain  asymmetry in the current-phase relation yielding non-zero JD efficiency  \cite{PhysRevX.12.041013,PhysRevLett.129.267702,matsuo2023josephson,zazunov2024nonreciprocal,PhysRevLett.130.266003,soori2023nonequilibrium,PhysRevB.109.094518,kwz7-5stj,PhysRevB.109.174511,mori2025towards,qzq2-f9kv,PhysRevResearch.5.033199,davydova2022universal,huamani2024theory,maiellaro2024engineered}. The electron and hole currents differ from each other under  IS-breaking while TRS-breaking causes  spin-up and spin-down electrons to carry different amount of current. The intrinsic JDE is analyzed in the absence of external magnetic field where finite momentum Cooper pair plays significant role \cite{PhysRevLett.128.037001,yuan2022supercurrent,he2022phenomenological}. The Andreev bound state (ABS) contribution to Josephson current is also studied in context of JDE \cite{PhysRevX.15.011046,79tj-c3y4}.  
Note that JDE is noticed in Rashba superconductor \cite{bauriedl2022supercurrent},  topological insulator \cite{Legg2022,PhysRevLett.131.096001}, Dirac semimetal \cite{pal2022josephson,PhysRevB.109.064511}, topological superconductors \cite{PhysRevB.109.L081405,PhysRevB.110.014519}.
Therefore, SDE and JDE both require MCA while they may have intriguing connection with the topological aspect of the systems.

The magnetism is another example of many-body order which is significantly different from superconductivity. A new magnetic order, altermagnetism (AM) \cite{Smejkal2022}, distinct from ferromagnetism or anti-ferromagnetism has been recently discovered experimentally \cite{PhysRevX.12.031042,reimers2024direct,ding2024large,yang2025three}. {AM is understood to originate from the non-relativistic spin splitting having even parity in momentum, whereas relativistic SOC such as Ising and Rashba SOCs are odd in momentum.} AM, breaking TRS without any net magnetization, is an admixture of ferromagnetic and antiferromagnetic order where the  Fermi-surface hosts  spin-split as well spin-degenerate features simultaneously.  
There have been a plethora of theoretical studies focusing on this unconventional $d$-wave magnetism \cite{tamang2025altermagnetism,cheong2025altermagnetism,PhysRevX.12.040002,tamang2024newly,cheong2025altermagnetism,PhysRevX.12.031042,PhysRevX.12.040501,PhysRevX.14.011019,cheong2024altermagnetism,PhysRevB.109.224502,fu2025all}. The real space rotation combined with translation symmetry protects the spin configuration of AM.  Interestingly, AM can naturally replace the role of the Zeeman field and therefore,  AM
is found to be very instrumental in the context of superconductivity  including  topological superconductivity \cite{4318-ttvf,PhysRevB.108.184505,PhysRevLett.133.106601,zylh-rqxl,PhysRevB.110.094508,PhysRevB.108.L060508,PhysRevLett.133.226002}, 
Weyl semimetal \cite{Li2025} and other electronic topological systems \cite{PhysRevB.111.184437,PhysRevB.111.085127,PhysRevB.111.155303}.

Given the above background, the AM is an immensely important material's property in the context of  SDE and JDE. The TRS breaking can be achieved by AM without the application of any external magnetic field.  This opens up an exciting field of  diode effects namely, field-free SDE \cite{dong2025field,li2024field,PhysRevLett.134.236703,samanta2025field,qi2025high,PhysRevB.110.024503,b7rh-v7nq,Fukaya_2025} and field-free JDE \cite{qzq2-f9kv,PhysRevB.110.014518,wu2022field,PhysRevB.110.104518,wu2021realization,PhysRevApplied.21.054057,debnath2025field}.  
The effect of crystallographic angle of the altermagnetic term is  also examined  to study JDE  in the presence of $p$-wave superconductivity in two-dimension (2D) \cite{Sharma2025}. The $p$-wave superconductivity is found to be essential to obtain JDE in one-dimension (1D) as well \cite{Soori2024}. On the other hand, two magnetic fields are required for  SDE to experience in 1D \cite{Legg2022,PhysRevB.107.224518} while AM induces field-free SDE in 2D \cite{PhysRevLett.134.236703,bhowmik2025field}. Interestingly, exploiting the MCA, one can also investigate the phenomenon of  SDE-mediated Josephson transistor  \cite{PhysRevB.108.174516}.  Although significant progress has been made in understanding SDE and JDE individually, their connection with FF pairing and MCA in field-free SDE remains largely unexplored. Motivated by this, we systematically explore the answers of the following questions by studying field-free SDE and JDE  in  1D: Can AM induce SDE in the absence of SOC (Ising/Rashba or both)?  What is the role of crystallographic angle in SDE and JDE as such can it alter the sign of the efficiency? How JDE is manifested in the FF superconductor where $p$-wave superconductivity is absent? Can JDE persist in the absence of AM and SOC?

In this paper, we uncover the field-free route to SDE and JDE by using AM, assembled with crystallographic angle $\alpha$, and two types of SOC such that MCA is present in the system,  {see Figs. (\ref{fig: Schematic}) (a,b,c) for more details on the setups and models. The parameter $\alpha$ denotes the crystallographic orientation angle associated with the AM order, see Fig. (\ref{fig: Schematic}) (d). In other words, $\alpha$ denotes an offset present in the AM term  which produces 
two-component AM order with $\sin k$ and $\cos k$ contributions. }
We first start with FF superconductor in the presence of AM, and Ising (Rashba) SOC in Model I (II) to study the SDE. Examining the low-energy BdG Hamiltonian, we find that the indirect band-gap closing with different probe current is connected with the non-reciprocity of the supercurrent with the finite momentum of the FF superconductor. While there exists qualitative similarity in the non-reciprocal behavior of supercurrent, the nature of SOC non-trivially governs the efficiency profile along with the strength and angle  of AM. Most importantly, two-component AM with finite $\alpha$ can generate MCA leading to an induction of  SDE even without any SOC in both the models. To be precise, the SDE is mainly driven by parallel spin components with complementary momentum functions in the presence of FF pairing. We tabulate the main findings of SDE in Table \ref{tab:novelty_intro_SDE}. 
We continue investigating the role of MCA in JDE where consider  $p$-wave and FF superconductivity in setup 1 and 2,3, respectively, along with AM and SOC in the intermediate quantum wire (FF superconductor itself) for setup 1,2 (3).  Our finding shows that bulk- (ABS-) contributions to Josephson current is significant to produce non-reciprocal current phase relation in setup 1 (3) while JDE is negligible in setup 2. Beside this,
sign of efficiency can be altered (remains unaltered) by the angle of AM  in setup 1 (2,3) where $p$-wave (FF) superconductivity is present. The field-free JDE is  more efficient in the presence of   $p$-wave superconductivity, AM and SOC. Similar to SDE, parallel spin components with complementary momentum functions with $p$-wave superconductivity  can yield JDE. 
Remarkably, the finite momentum in FF superconductivity breaks IS and TRS, resulting in non-zero JD efficiency via MCA even in the absence of AM, SOC and $p$-wave superconductivity. We tabulate the main findings of JDE in Table \ref{tab:novelty_intro_JDE}.

\begin{widetext}

\begin{table}[H]
\begin{tabular}{| p{2cm}|p{2cm}|p{2cm}|p{2cm}|p{2cm}|p{2cm} |p{2cm}| p{2cm}|} 

\hline External
 Super-current  &  AM strength  & AM angle &  Rashba SOC  & Ising SOC &  MCA & Emergence of FF state ($j(q_0)=0$ at $q_0\ne 0$) & Field-free SDE in 1D \\ 
 
\hline 
$q \ne 0$   & $J_A=0$ &  $\alpha=0$ or $\ne0$ &  Absent   &  Absent & Present as IS and TRS are broken & Not possible & Absent  \\ 
\hline
$q \ne 0$   & $J_A \ne 0$ &  $\alpha=0$ &  Absent   &  Present & Present as IS and TRS are broken & Possible & Possible as shown in Model I  \\ 
\hline
$q \ne 0$   & $J_A \ne 0$ &  $\alpha=0$ &  Present   &  Absent & Present as IS and TRS are broken & Possible & Absent as shown in Model II  \\ 
\hline
$q \ne 0$   & $J_A \ne 0$ &  $\alpha \ne 0$ &  Absent   &  Absent & Present as IS and TRS are broken & Possible & Possible due to parallel spin components  \\ 
\hline
\end{tabular}
\caption{We tabulate the SDE for different models along with their ingredients. Note that in the absence of both SOCs, only two-component altermagnetism with finite crystallographic angle $\alpha\ne 0$, could lead to SDE, see the last row. }
\label{tab:novelty_intro_SDE}
\end{table}


\begin{table}[H]

\begin{tabular}{| p{2cm}|p{2cm}|p{2cm}|p{2cm}|p{2cm}|p{2cm} |p{2cm}| p{2cm}|} 
\hline Type of superconductivity  &  AM strength irrespective of $\alpha$  &  Rashba SOC  & Ising SOC & External supercurrent &  Quasi contribution to diode effect & ABS contribution to diode effect  & Field-free JDE in 1D \\ 
\hline
 $p$-type  & $J_A=0$ in quantum wire & Absent in quantum wire  &  Absent in quantum wire  &  Absent & Absent  & Absent & Absent   \\ 
 \hline
 $p$-type  & $J_A \ne 0$ in quantum wire & Absent  or present in quantum wire  &  Absent or present in quantum wire  &  Absent & Present   & Present & Present with higher efficiency  as found in setup 1\\ 
 \hline
 $s$-type  & $J_A= 0$ or $\ne 0$ in quantum wire & Absent or present in quantum wire  &  Absent or present in quantum wire &  Present & Minimally present  & Maximally present & Present with lower efficiency  as found in setup 2  \\ 
 \hline
 $s$-type  & $J_A= 0$ or $\ne 0$  in superconductor & Absent or present in superconductor  &  Absent or present in superconductor &  Present & Minimally present  & Maximally present & Present with intermediate efficiency as found in setup 3 \\ 
\hline 

\end{tabular}
\caption{We tabulate the JDE for different setups along with their ingredients. We find that JDE can be solely mediated by finite supercurrent even in the absence of $p$-wave pairing, SOCs, and altermagnetism, see setups 2 and 3.  }
\label{tab:novelty_intro_JDE}
\end{table}

\end{widetext}

The paper is organized as follows: In Sec. \ref{sec:sde}, we discuss two different SDE models, methods to compute supercurrent, and diode efficiency, and findings associated with these models. In Sec. \ref{sec:jde}, we demonstrate three setups for Josephson junction, formalism to compute the Josephson current, and diode efficiency, and findings corresponding to these setups. In Sec. \ref{connection}, we connect the field-free SDE and JDE in terms of the sources of MCA.  In Sec. \ref{conclusion}, we conclude with future directions.

\section{Superconducting Diode Effect (SDE)}
\label{sec:sde}

\subsection{Model and band dispersion}
\label{ssec:sde1}

We consider a simple 1D tight-binding model with IS breaking Rashba (Ising) SOC and TRS breaking  AM of strengths $\lambda_R (\lambda_I)$ and $J_A$, respectively \cite{PhysRevX.12.040501}. {For improved clarity, we have shown a schematic of the setup used in this analysis in Fig. \ref{fig: Schematic} (a).} The normal-state with spinful lattice Hamiltonian  is then given by $H_0 (\lambda_{I},\lambda_{R})= \frac{1}{2}\sum_{n=1}^{L-1}  \bigg[ t \sum_{\sigma=\uparrow,\downarrow} c_{n,\sigma}^\dagger c_{n+1,\sigma} -   i \lambda_{I} \big( c_{n,\uparrow}^\dagger c_{n+1,\uparrow} - c_{n,\downarrow}^\dagger c_{n+1,\downarrow}\big) -  i \lambda_{R}\big( c_{n,\uparrow}^\dagger c_{n+1,\downarrow} + c_{n,\downarrow}^\dagger c_{n+1,\uparrow}\big) +  J_A \big[ \cos 2\alpha (c_{n,\uparrow}^\dagger c_{n+1,\uparrow} -c_{n,\downarrow}^\dagger c_{n+1,\downarrow})   
+ i \sin 2\alpha (c_{n,\uparrow}^\dagger c_{n+1,\uparrow} -c_{n,\downarrow}^\dagger c_{n+1,\downarrow}) \big] +{\rm h.c.} \bigg ]= \sum_{n,m}\tilde{\Psi}_n^\dagger \tilde{H}^{nm}_{0}(\lambda_{I},\lambda_{R})\, \tilde{\Psi}_m$ where $L$ is the number of lattice sites, $t $ represents the nearest-neighbour hopping. and $\tilde{\Psi}_m=(c_{m,\uparrow},c_{m,\downarrow})^T$.  Here $\alpha$ is the crystallographic angle of the AM \cite{Sharma2025}.  Now coming to the superconducting part,  the FF pairing is taken as on-site $(\Delta(n) c_{n,\uparrow}^\dagger c_{n,\downarrow}^\dagger + {\rm h.c.})$  with a spatially varying order parameter for the pair-density wave,
$\Delta(n) = \Delta\, e^{i q n}$ where $n = 1,\dots,L$ denotes the site-index.   Finally, assembling particle and hole blocks  in 
the Nambu basis $\Psi_n^\dagger = (\tilde{\Psi}^{\dagger}_n,\tilde{\Psi}_n)=(c_{n\uparrow}^\dagger,\, c_{n\downarrow}^\dagger,\, c_{n\uparrow},\, c_{n\downarrow})$ on $n$-th site, one obtains the lattice BdG Hamiltonian
\begin{equation}
H^{\rm I(II)}_{\mathrm{BdG}}(q) = 
\begin{pmatrix}
\tilde{H}^{\rm I(II)}_0{(\lambda_I,\lambda_R)} -\mu I & \widehat{\Delta} \\
\widehat{\Delta}^\dagger & -\big[\tilde{H}^{\rm I(II)}_0{(\lambda_I,\lambda_R)}\big]^\ast +\mu I
\end{pmatrix},
\label{eq:1}
\end{equation}
which is a \(4L\times 4L\) Hermitian matrix. {$\tilde{H}^{\rm I(II)}_0(\lambda_I,\lambda_R)$ represents the normal part of the Hamiltonian for model I or model II  and they depend on the Ising SOC strength $\lambda_I$ and Rashba SOC strength $\lambda_R$.} Note that the pairing off-diagonal block in spin space is $
\widehat{\Delta} = -\,i\sigma_y \otimes \Delta(n)$  where  the diagonal matrix $\Delta(n) = \Delta ~\mathrm{diag}\{e^{-i q n}\}$ where $\Delta$ denotes the  momentum independent superconducting gap amplitude. Here, $\mu I \equiv - \mu \sum_{n,\sigma} c_{n,\sigma}^\dagger c_{n,\sigma}$ with $\mu$ being the chemical potential. For our analysis, we take into consideration $\lambda_I \ne 0=\lambda,\lambda_R= 0$ [$\lambda_I =0 ,\lambda_R \ne  0=\lambda$] in $\tilde{H}^{\rm I}_0(\lambda,0)$ [$\tilde{H}^{\rm II}_0(0,\lambda)$] to construct Model I (II) with BDG Hamiltonian $H^{\rm I}_{\mathrm{BdG}}(q)$ ($H^{\rm II}_{\mathrm{BdG}}(q)$), { see Fig. \ref{fig: Schematic} (a).}

One can Fourier transform the above BdG Hamiltonian Eq. (\ref{eq:1}) to obtain the momentum space BdG  Hamiltonian in the Nambu  basis $\Psi_k=\Big(c_{k+\tfrac{1}{2}q,\uparrow},\;
c_{k+\tfrac{1}{2}q,\downarrow},\;
c^{\dagger}_{-k+\tfrac{1}{2}q,\uparrow},\;
c^{\dagger}_{-k+\tfrac{1}{2}q,\downarrow}\Big)^T$, given by
\begin{equation}
\mathcal{H}^{\rm I(II)}_k(q,\Delta) =
\begin{pmatrix}
H_{\rm I(II)}(k+\tfrac{1}{2}q) -\mu  & -i\sigma_y \Delta \\
i\sigma_y \Delta^* & -H_{\rm I(II)}^{*}(-k+\tfrac{1}{2}q) +\mu
\end{pmatrix}.
\label{eq:2}
\end{equation}
The diagonal kinetic parts {$H_{\rm I}(k)$, and $H_{\rm II}(k)$, which are the Fourier transforms of $\tilde{H}^{\rm I}_0(\lambda, 0)$, and $\tilde{H}^{\rm II}_0(0,\lambda)$, respectively,} depends on the momentum $q$ of the FF pairing and off-diagonal superconductivity is independent of $q$. In general, the kinetic part is given by 
\begin{align}
H_{\rm I(II)}(k) &= t \cos k  \,\sigma_0
+  \lambda \sin k\,\sigma_\beta  \notag \\
&\quad + J_A \big( \cos k\cos 2\alpha
+ \sin k \sin 2\alpha \big) \sigma_z .
\label{eq:3}
\end{align}
where $\lambda$ ($J_A$) identifies the strength of SOC (AM). 
Here, $\sigma_{x,y,z}$ are the Pauli matrices in spin space, and $\sigma_0$ is the $2\times 2$ identity.     Importantly, the AM term can be thought of as a dressed AM term $J_A \cos (k- 2 \alpha)$ which reduces to conventional AM term in the absence of the  crystallographic angle. Finite $\alpha$ produces two-component AM. 
Note that $\beta=x$ represents the Rashba SOC while $\beta=z$ denotes the Ising SOC.  Therefore, we treat Ising case as Model I with BdG Hamiltonian $\mathcal{H}_k^{\rm I}(q,\Delta)$ and Rashba case as Model II with BdG Hamiltonian $\mathcal{H}_k^{\rm II}(q,\Delta)$, having kinetic terms as   $H_{\rm I}(k)$ and $H_{\rm II}(k)$, respectively. {A schematic representation of the two models are shown in Fig. \ref{fig: Schematic} (b).}  We consider hopping $t$ and  other parameters in the units of meV. {The choice of energy units in meV is motivated by typical experimental energy scales in condensed matter systems, particularly in superconducting materials and nanoscale heterostructures \cite{Ando2020, bauriedl2022supercurrent, pal2022josephson, turini2022josephson, trahms2023diode, li2024field, zhou2025superconducting}.}

{
In standard two-dimensional (2D) $d$-wave altermagnets, the spin splitting is purely momentum-even. In other words,  altermagnetism is generically a spatial 2D phenomenon.  The 1D analogue of altermagnetism can be understood in terms of a heterostruture where the underlying 2D altermagnet can induce altermagnetism via a proximity effect on  1D chain embedded in the 2D sheet. In particular, we have a 1D superconductor oriented at an angle $\alpha$ with respect to the principal axes of the 2D altermagnetic substrate to induce the AM, which scales as $J_A \cos(k - 2\alpha)\sigma_z$ \cite{amartya2025, PhysRevLett.132.036702}. However, when a 1D nanowire is oriented along an arbitrary crystallographic axis at angle $\alpha$ relative to the principal axes of the momentum-even 2D altermagnetic substrate, the effective 1D subband projection acquires both $\cos k$ (even-parity) and $\sin k$ (odd-parity) projections along the wire direction. Under spatial inversion along the 1D wire, the term $(\cos k \cos 2\alpha) \sigma_z$ is even under inversion ($k$-even), whereas $(\sin k \sin 2\alpha) \sigma_z$ changes sign and is therefore explicitly inversion-breaking ($k$-odd).}

\begin{figure}
    \centering
    \includegraphics[width=0.5\textwidth]{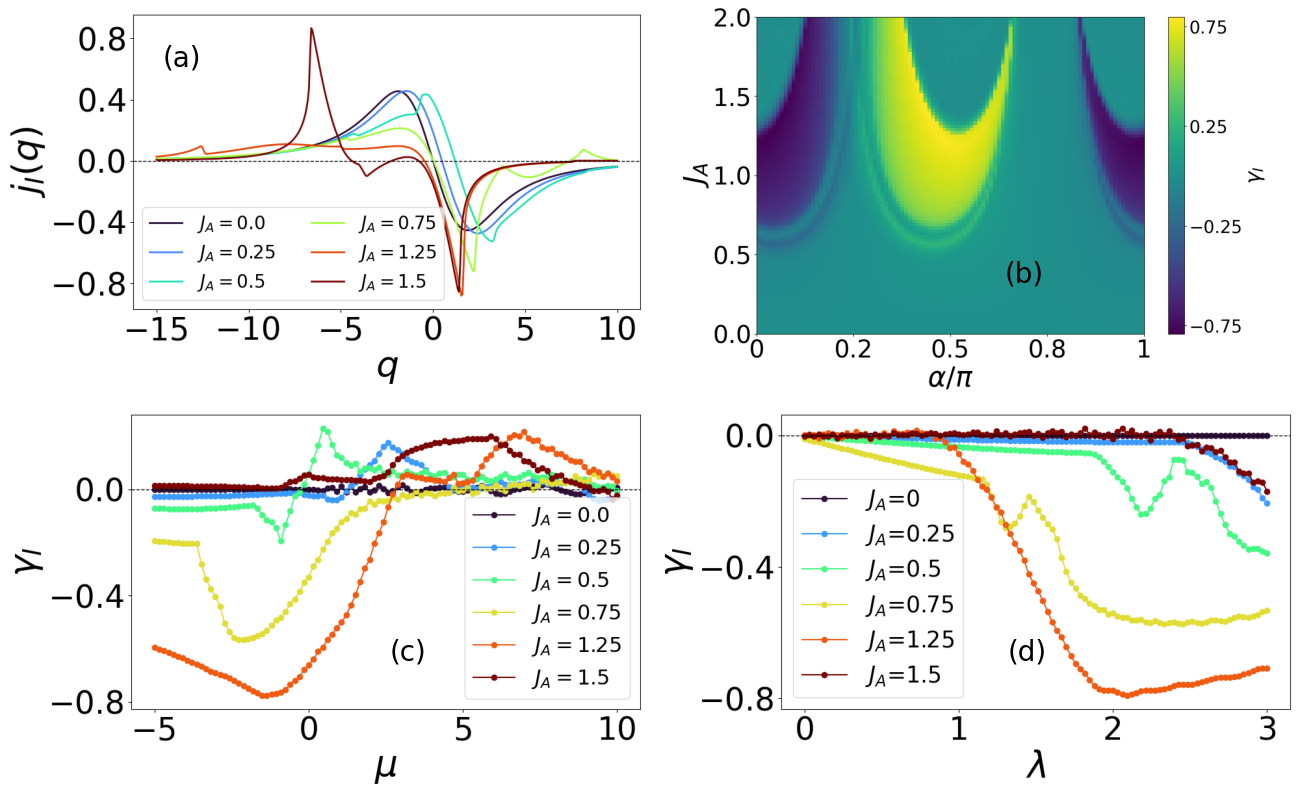}
    \caption{In (a), we show the variation of supercurrent $j_{\rm I}(q)$, derived from Eqs. (\ref{eq:5}), (\ref{eq:6}) for Model I low-energy Hamiltonian, with the finite momentum of the Cooper pair. In (b), we illustrate the heat-map of $\gamma_{\rm I}$ derived from Eq. (\ref{eq:7}) as a function of angle $\alpha$, and amplitude $J_A$ of altermagnet term. In (c,d), we display the variation of SD efficiency with $\mu$ and Ising SOC $\lambda$, respectively. We consider $\mu=-1.52$ for (a),(b) and (d). We choose $\lambda=1.97$ for (a),(b) and (c). We choose $\alpha=0.017\pi$ for (a),(c) and (d).
    Parameters: $t=1$, $\Delta=2.02$, $\beta=10000$.}
    \label{fig:5}
\end{figure}

\begin{figure}
    \centering
    \includegraphics[width=0.5\textwidth]{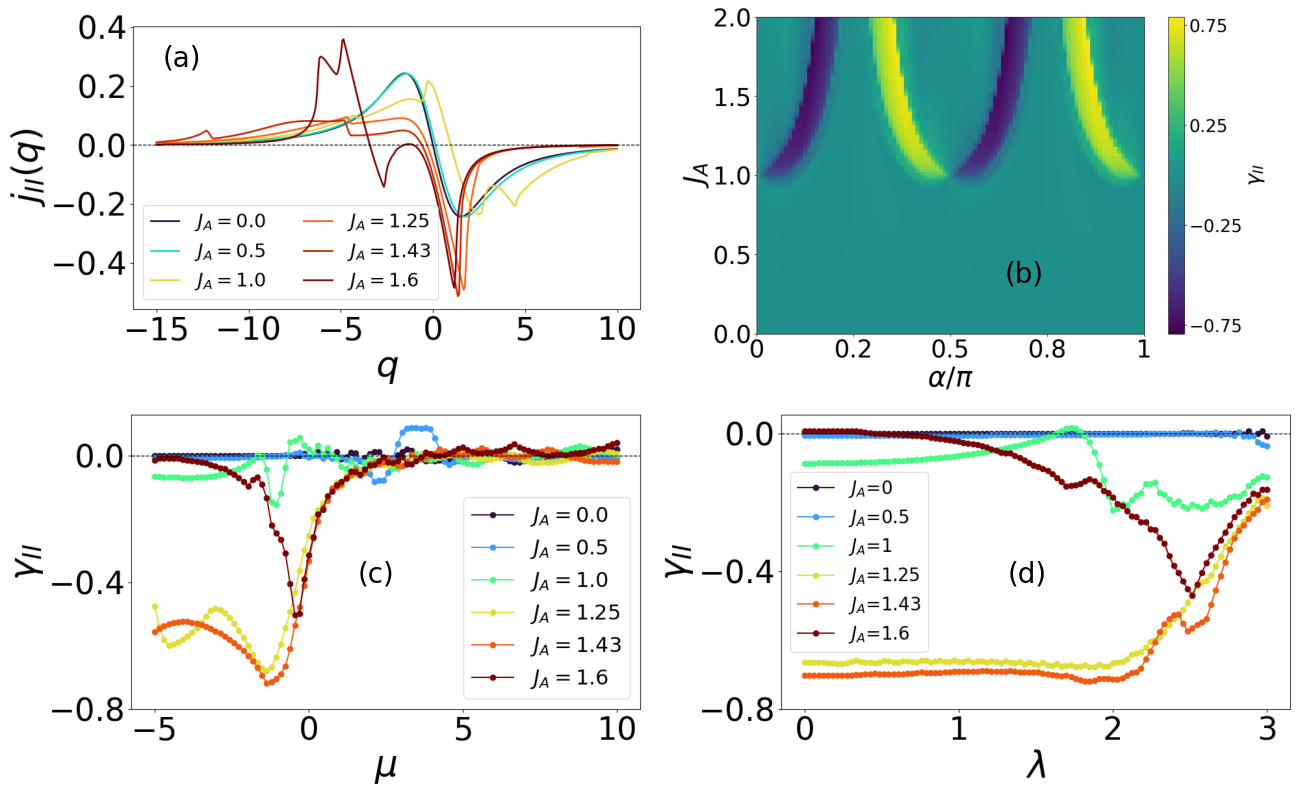}
    \caption{In (a), we show the variation of supercurrent $j_{\rm II}(q)$, derived from Eqs. (\ref{eq:5}), (\ref{eq:6}) for Model II low-energy Hamiltonian, with the finite momentum of the Cooper pair. In (b), we illustrate the heat-map of $\gamma_{\rm II}$ derived from Eq. (\ref{eq:7}) as a function of angle $\alpha$, and amplitude $J_A$ of altermagnet term. In (c,d), we display the variation of SD efficiency with $\mu$ and Rashba SOC $\lambda$, respectively. We consider $\mu=-1.36$ for (a),(b) and (d). We choose $\lambda=1.87$ for (a),(b) and (c). We choose $\alpha=0.111\pi$ for (a),(c) and (d).
    Parameters: $t=1$, $\Delta=1.21$, $\beta=10000$.}
    \label{fig:6}
\end{figure}

In the absence of SOC i.e., $\lambda=0$ and $\alpha=0$, the normal part $H_{\rm I(II)}(k)$ preserves  IS, generated by $\mathcal {C}=\sigma_z$, such that $\mathcal {C} H_{\rm I(II)}(k) \mathcal {C}^{-1}=H_{\rm I(II)}(-k)$. In the absence of AM i.e., $J_A=0$,  $H_{\rm I(II)}(k)$ preserves  TRS $\mathcal {T}= \mathcal {K} \sigma_y$ such that $\mathcal {T} H_{\rm I(II)}(k) \mathcal {T}^{-1}= H_{\rm I(II)}(-k)$, $\mathcal {K}$ denotes complex conjugation. We note that $H_{\rm I(II)}(k)$
breaks IS and TRS when in the presence of SOC $\lambda \ne 0$ and AM $J_A \ne 0$, respectively.  It is noteworthy that $\alpha \ne 0$ also breaks IS even when $\lambda=0$ when the kinetic Hamiltonian $H_{\rm I(II)}(k)$ only contains AM.  
The BdG Hamiltonian with $\lambda=0$, $\alpha=0$ and $q=0$ preserves IS such that $ \mathcal {\tilde C} \mathcal{H}^{\rm I(II)}_k(0,\Delta)  \mathcal {\tilde C}^{-1} = \mathcal{H}^{\rm I(II)}_{-k}(0,\Delta)$ with $\mathcal {\tilde C}= \mathcal {C} \tau_z$. Importantly, for $q \ne 0$, $\mathcal{H}^{\rm I(II)}_k(q,\Delta)$ no longer preserves IS
even if $\lambda=0$ and $\alpha=0$. 
Now coming to the TRS of BdG Hamiltonian with $J_A=0$ and $q=0$,  $\mathcal {\tilde T}= \mathcal {T} \tau_z$ such that $\mathcal {\tilde T} \mathcal{H}^{\rm I(II)}_k(0,\Delta)  \mathcal {\tilde T}^{-1} = \mathcal{H}^{\rm I(II)}_{-k}(0,\Delta)$. Importantly, for $q \ne 0$, $\mathcal{H}^{\rm I(II)}_k(q,\Delta)$  breaks TRS even if $J_A=0$.    This indicates the fact that finite momentum pairing causes IS and TRS breaking in the superconductor irrespective of SOC and AM. Therefore, finite momentum  $q\ne 0$,
is a sufficient condition for SDE to occur, while the breaking of IS and TRS in the normal part $H_{\rm I(II)}(k)$ is a necessary condition. It is important to note that SOC and magnetic field are required to generate finite momentum of the Cooper pair \cite{qu2013topological}.  In short, TRS and IS breaking are mandatory conditions to obtain SDE. In this work we have chosen to impose $q$ by hand, however one can determine $q_0$ by self-consistently minimising free energy $F(q, \Delta)$. {In practice, the magnitude of $q$ cannot be arbitrarily large. When $q$ becomes sufficiently large, it can significantly perturb the superconducting state and eventually suppress superconductivity \cite{PhysRevB.111.L161402,bhowmik2025optimizing}. Therefore, in our analysis $q$ is chosen within a physically reasonable range, typically $q \sim \mathcal{O}(t,\lambda,J_A,\Delta)$, ensuring that the superconducting state remains stable while still capturing the effects of finite-momentum pairing. This behavior is illustrated in Figs.~\ref{fig:5}(a) and \ref{fig:6}(a), as explained later,  where the supercurrent decreases and eventually approaches zero for large values of $q$. This suppression reflects the weakening and eventual destruction of superconductivity at large pairing momentum.}

As discussed earlier, the non-reciprocity in the critical current is the main essence of SDE. Upon changing the direction of magnetic field or probe current, the critical current also alters in magnitude in addition to its direction. 
It has been shown that magnetochiral anisotropy is required to obtain SDE   \cite{li2024field,bauriedl2022supercurrent,nadeem2023superconducting}. In dimension $D\ge 2$,  the critical current is given by $j_c(B)=j_c(0)[1+ \tilde{\gamma} ({\hat C} \times {\bar B}).{\hat J}]$ where ${\hat C}$ denotes the direction of IS breaking, ${\bar B}$ represents the magnetic field which can be in-plane (out-of-plane) if SOC is out-of-plane (in-plane) and ${\hat J}$ is the direction of the probe current. The probe current effectively mimics the finite momentum of the Cooper pair. $\tilde{\gamma}$ denotes the strength of magnetochiral anisotropy. Therefore, directions of IS breaking, magnetic field and Cooper pair momentum have to be orthogonal to each other to experience the effect of $\tilde{\gamma}$. The anisotropy in current thus can be examined either by reversing the magnetic field or the probe current such that $ {\bar B} \to - {\bar B}$  (${\hat J} \to -{\hat J}$) keeping ${\hat J}$ (${\bar B}$) unaltered, leading to $j^+_c \ne j^-_c$.  Note that the Pauli matrices associated with SOC and magnetic field are considered to be the direction of $\hat C$ and $\hat B$, respectively, in $D\ge 2$. Interestingly, for our study in  $D=1$, parallel spin matrices corresponding to IS and TRS breaking components contribute to MCA that we explore extensively while studying SDE and JDE.

Here, our aim is to demonstrate SDE in the absence of any external magnetic field, rather in the presence of AM. The AM is associated with $\sigma_z$ due to its local magnetic moments restricting us from altering its direction unlike the external magnetic field. Given the fact that AM is along the out-of-plane direction,  we allow for two different types of SOC, namely in-plane Rashba and out-of-plane Ising SOC, to examine SDE \cite{nadeem2023superconducting}. It is important to note that in-plane Rashba (out-of-plane Ising) SOC and out-of-plane (in-plane) magnetic field are responsible for observing SDE. 
We investigate the non-reciprocity in supercurrent by changing the sign of $q$ which is the same as changing the direction of the  probe current $J$.  Therefore, we investigate AM-mediated SDE following two routes namely,  Model I with Ising SOC and  Model II with Rashba SOC.

\begin{figure}
    \centering
    \includegraphics[width=0.5\textwidth]{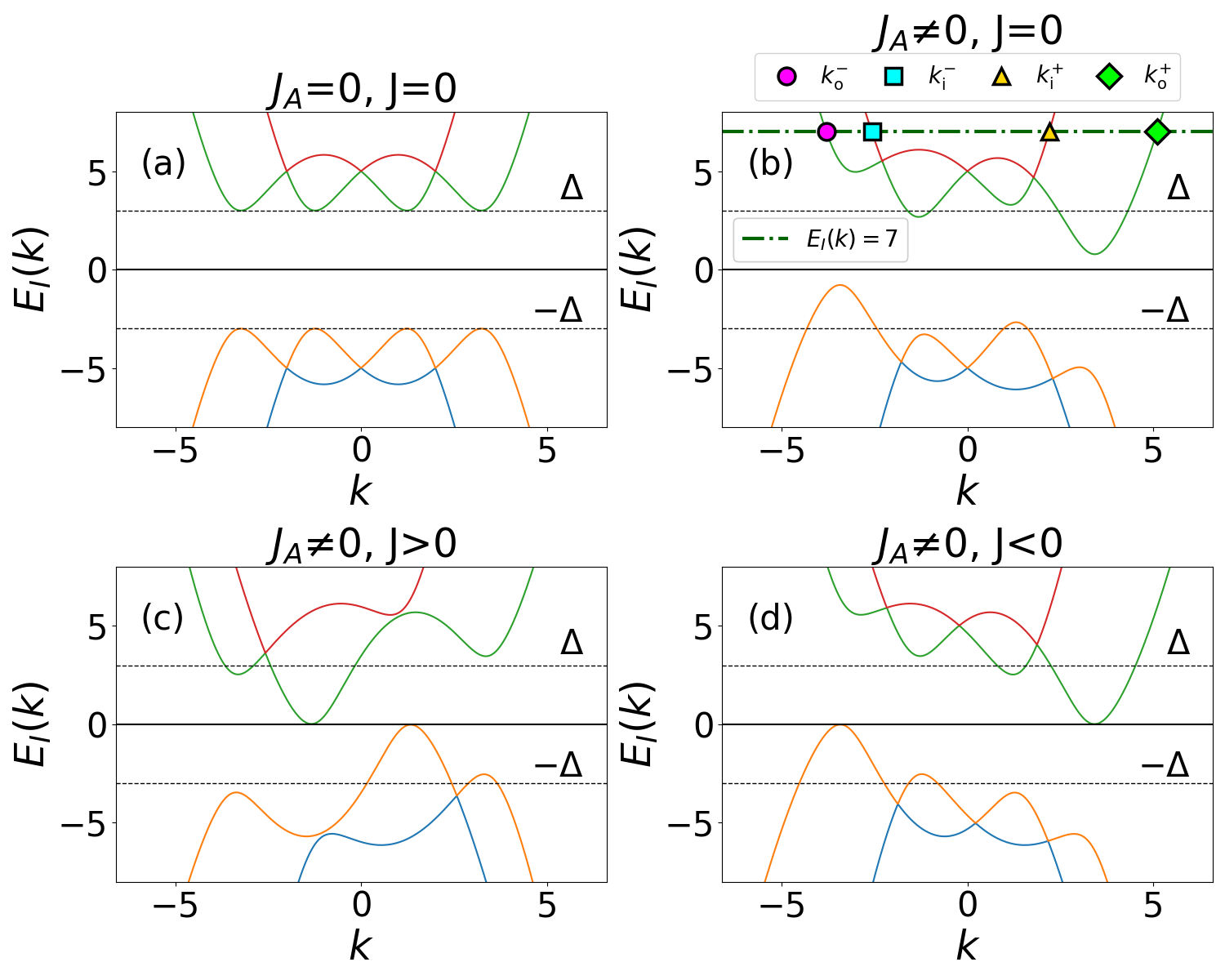}
    \caption{We show $E_{\rm I}(k)$, derived from Eq. (\ref{eq:2}) and (\ref{eq:4}) for Model I low-energy Hamiltonian, for $(J_A,J)=(0,0), (0.2,0), (0.2,1.18),$ and $(0.2,-0.35)$ in (a), (b), (c), and (d), respectively. The plot (b) depicts the inner and outer circle crossings for $E_{\rm I}(K) = 7$. For this Ising Case: $k_o^-=-3.796$, $k_i^-=-2.572$, $k_i^+=+2.199$, $k_o^+=+5.103$. Parameters: $t=1$, $\mu=4$, $\Delta=3$, $\lambda=2$, $\alpha=0$.}
    \label{fig:1}
\end{figure}

\begin{figure}
\includegraphics[width=0.48\textwidth]{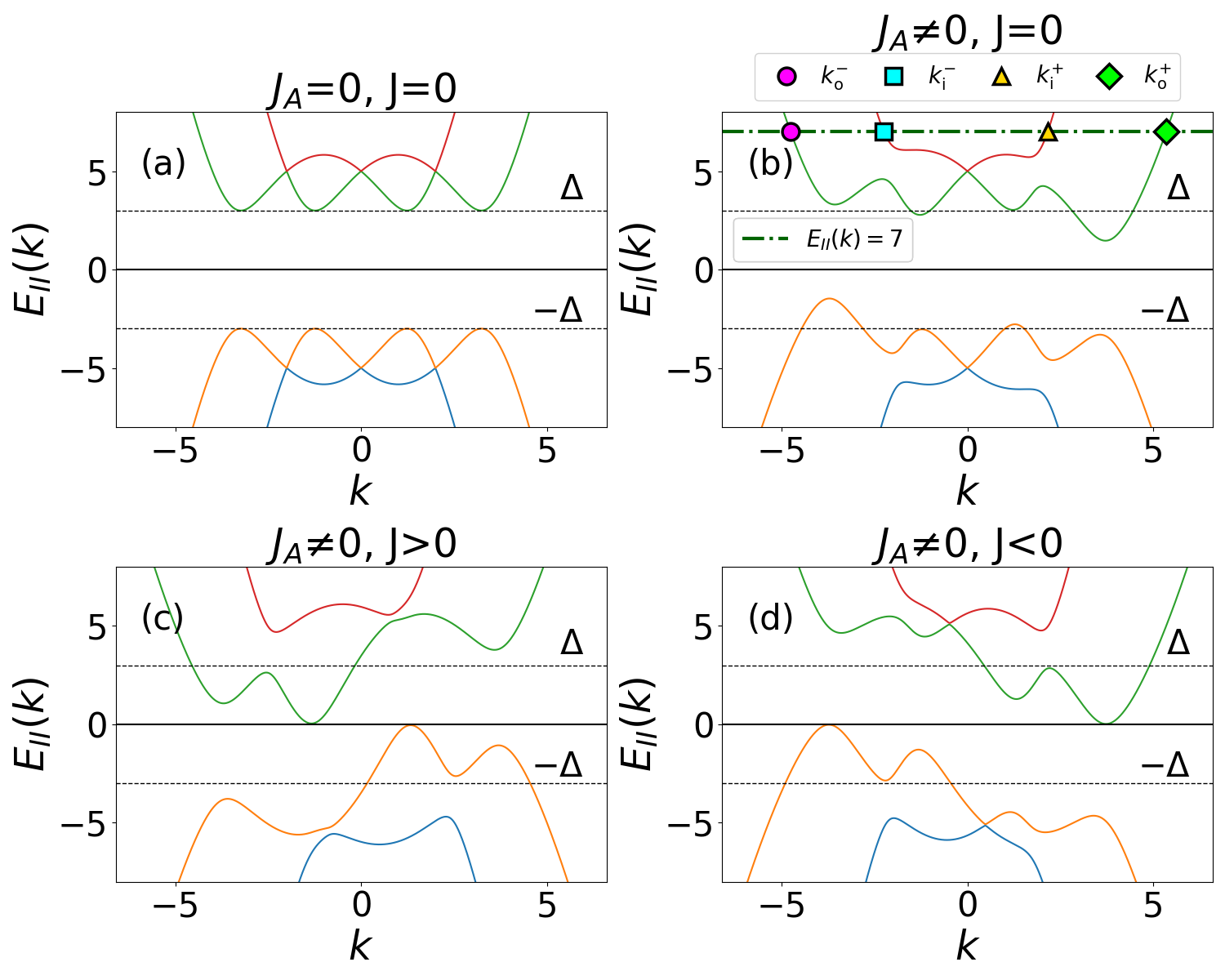}
\caption{We show $E_{\rm II}(k)$, derived from Eq. (\ref{eq:2}) and (\ref{eq:4}) for Model II low-energy Hamiltonian, for $(J_A,J)=(0,0), (0.6,0), (0.6,1.18),$ and $(0.6,-0.75)$ in (a), (b), (c), and (d), respectively. The plot (b) depicts the inner and outer circle crossings for $E_{\rm II}(K) = 7$. For this Rashba Case: $k_o^-=-4.757$, $k_i^-=-2.247$, $k_i^+=+2.159$, $k_o^+=+5.347$. Parameters: $t=1$, $\mu=4$, $\Delta=3$, $\lambda=2$, $\alpha=0.125\pi$.}
\label{fig:2}
\end{figure}


We now present the low-energy BdG model with $\cos k \approx k^2$ {neglecting the constant term without the loss of generality \cite{Legg2022}}, and $\sin k \approx k$ in  Eq.(\ref{eq:2}) yielding the normal part   
\begin{align}
H_{\rm I(II)} (k) & \approx t k^2 \,\sigma_0
+  \lambda  k\sigma_{z(x)}  \notag \\
&\quad + J_A \big( \ k^2 \cos 2\alpha
+  k \sin 2\alpha \big) \sigma_z .
\label{eq:4}
\end{align}
The corresponding BdG Hamiltonian Eq. (\ref{eq:2}) is given by  $\mathcal{H}^{\rm I(II)}_k(q,\Delta) \approx \big(t k^2 - \mu\big)\sigma_0\tau_z +  \lambda  k\sigma_{z(x)} \tau_0 +J_A k^2 \cos (2\alpha) \sigma_z \tau_z +J_A k \sin (2\alpha) \sigma_z \tau_0 + t (q^2/4) \sigma_0\tau_z + t kq \sigma_0\tau_0 + \lambda (q/2) \sigma_{z(x)} \tau_z +J_A (q^2/4) \cos (2\alpha) \sigma_z\tau_z + J_A kq \cos (2 \alpha) \sigma_z \tau_0 + (q/2) \sin (2 \alpha) \sigma_z \tau_z + \Delta \sigma_y \tau_y $. 

\begin{figure}
\centering
\includegraphics[width=0.48\textwidth]{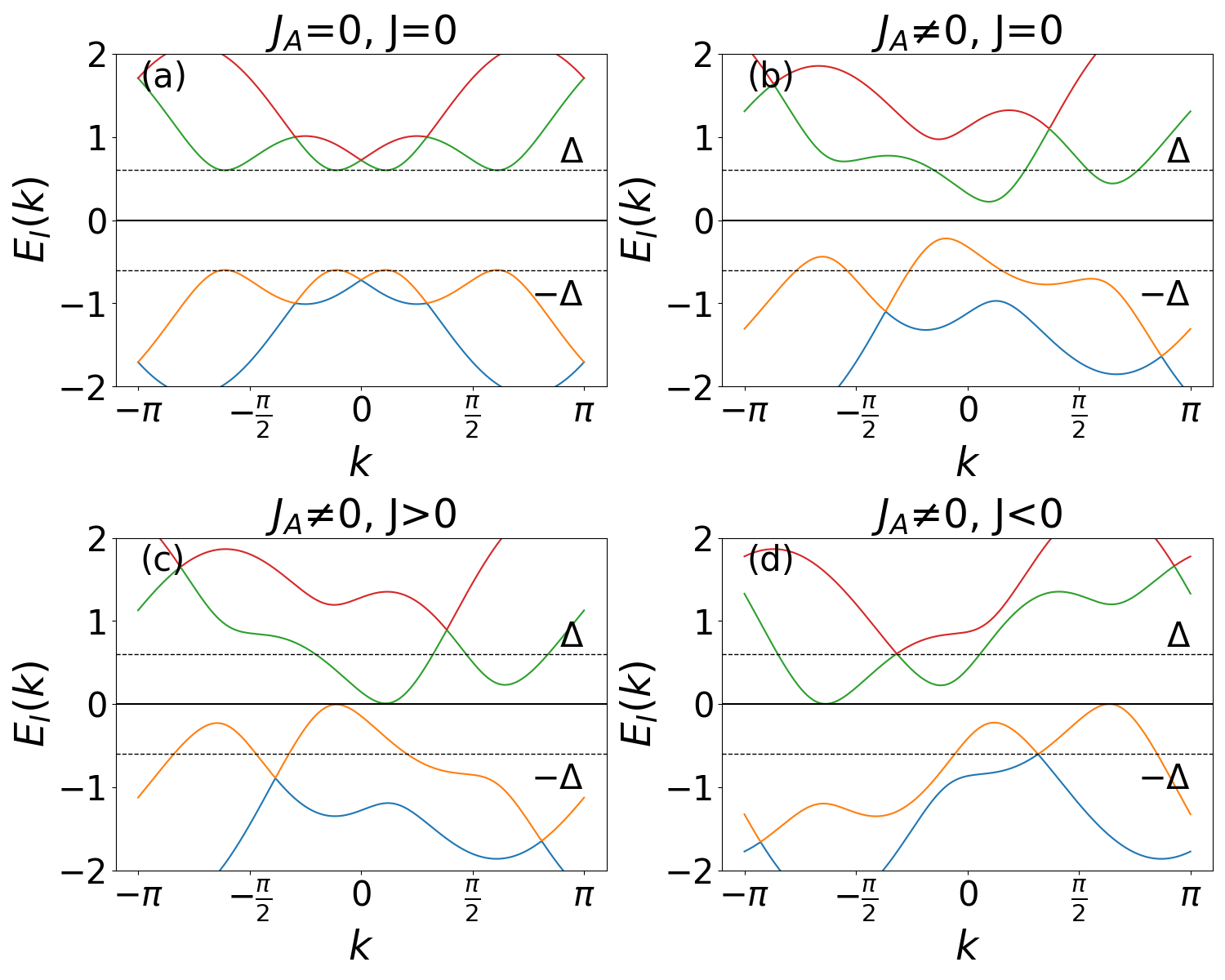}
\caption{We show $E_{\rm I}(k)$, derived from Eq. (\ref{eq:2}) and (\ref{eq:3}) for Model I \textcolor{red}{tight-binding} Hamiltonian, for $(J_A,J)=(0,0), (0.4,0), (0.4,0.35),$ and $(0.4,-1.18)$ in (a), (b), (c), and (d), respectively. Parameters: $t=1$, $\mu=0.6$, $\Delta=0.6$, $\lambda=1$, $\alpha=0$.}
\label{fig:3}
\end{figure}


\begin{figure}
\includegraphics[width=0.48\textwidth]{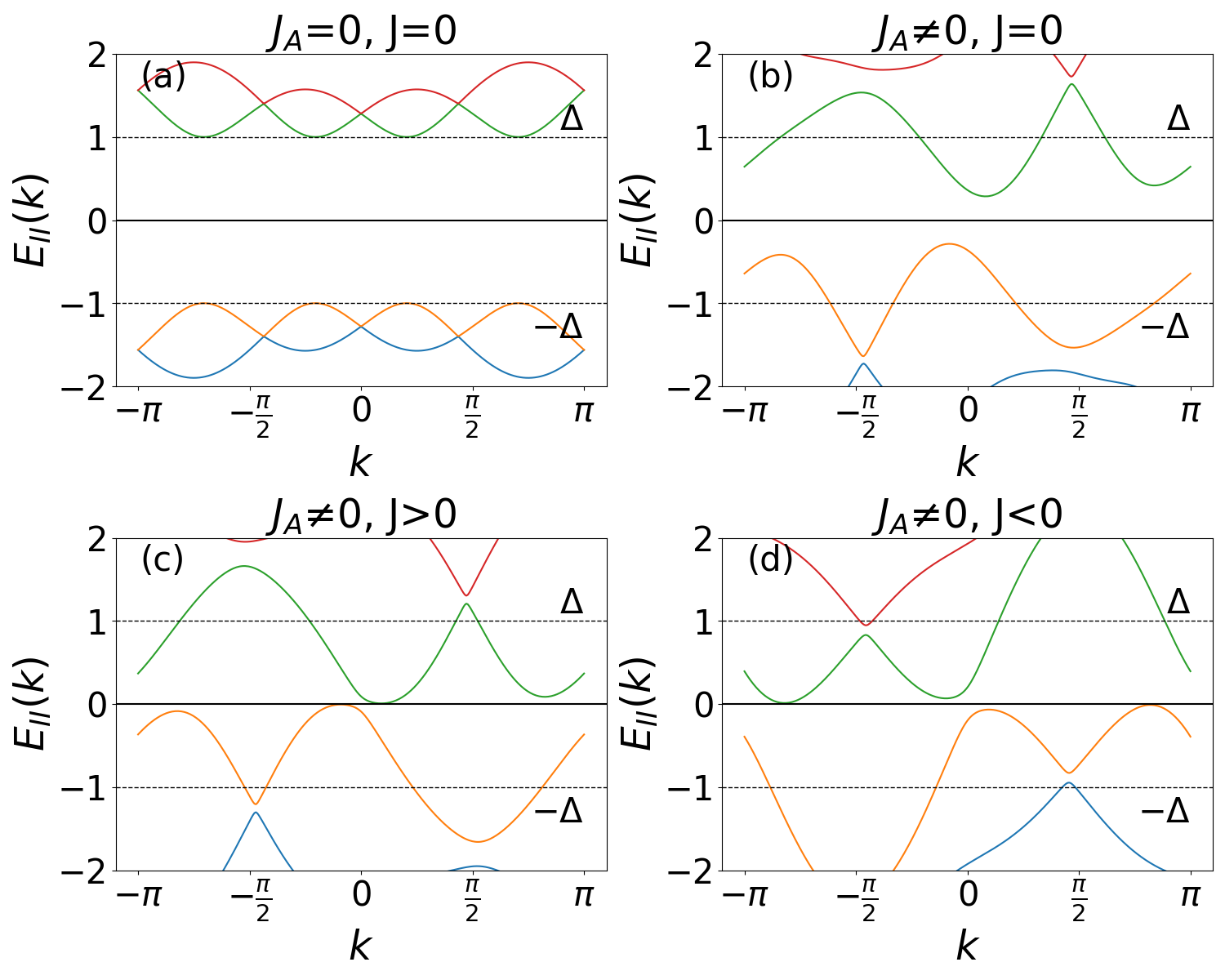}
\caption{We show $E_{\rm II}(k)$, derived from Eq. (\ref{eq:2}) and (\ref{eq:3}) for Model II {tight-binding} Hamiltonian, for $(J_A,J)=(0,0), (1.3,0), (1.3,0.55),$ and $(1.3,-1.95)$ in (a), (b), (c), and (d), respectively. Parameters: $t=1$, $\mu=0.2$, $\Delta=1$, $\lambda=1$, $\alpha=0.125\pi$.}
\label{fig:4}
\end{figure}


We now examine the band dispersion from the low-energy model in  Fig. \ref{fig:1} for the Model I with Ising SOC. The Ising SOC lifts the degeneracy of the earlier quadratic spin-degenerate bands except at $k=0$. 
In the absence of  $J_A$, see Fig. \ref{fig:1} (a), the isotropic nature of the bands is observed with degeneracy at $k=0$.
This results in concentric Fermi surfaces when the Fermi energy is kept above $E_k>5$. The center of the inner and outer rings overlap at $k=0$. In the presence of finite $J_A \ne 0$, the Fermi surfaces no longer remain concentric leading to a separation between the centers of inner and outer rings, see Fig. \ref{fig:1} (b). In other words, $J_A$ introduces anisotropy in the dispersion which could potentially lead to coupling between two electrons of opposite  spins and opposite but unequal momentum $k^+_{o,i} \ne k^-_{o,i}$. In the presence of AM, the Fermi surfaces are deformed from their circular structure. 
Therefore, the deviation of the centers $k^c_{o,i}=(k^+_{o,i} + k^-_{o,i})/2$ of inner and outer rings  with respect to $k=0$ can effectively cause a finite Cooper pair momentum $q_0$ to emerge. $k^c_{o}$ and $k^c_{i}$ are of opposite signs.  It turns out that the change in the finite momentum, i.e., $\delta q \equiv q - q_0$, due to the passing of a current through the superconductor is proportional to the current  \cite{yuan2022supercurrent}. Interestingly, the probe current $J$  directly 
accounts for $q$ i.e., $J \equiv q$ in the BdG Hamiltonian and hence passing a probing current in the direction of (against) $q_0$ eventually closes the gap of the inner (outer) pocket, see Figs. \ref{fig:1} (c,d). 
This gap closing in opposite directions happens at different values of $J$ namely $J_{\pm}$ which signifies the asymmetric nature of the indirect band gap as a function of $q$ with respect to $q=0$. 
This directly indicates that the  superconducting gap vanishes asymmetrically with $J \equiv q$ causing  different critical currents in opposite directions. Therefore,  the superconducting diode effect takes place as long as  $J_+\neq J_-$ or $q_+\neq q_-$ features are evident from band dispersion \cite{Fu2022, Legg2022}.

We examine the dispersion of the low-energy version of Model II with Rashba SOC in Figs. \ref{fig:2} (a,b,c,d) where we qualitatively find identical features as those of Model I. In the presence of $J_A$, the inner and outer Fermi surfaces no longer remain concentric. Under positive and negative probe currents, the  band gaps of inner  and outer surfaces close at different values of $J$ suggesting the emergence of SDE. To take this a step further, we study the dispersion of the  tight-binding version of Model I and II in Figs. \ref{fig:3} and \ref{fig:4}, respectively. The asymmetric behavior of the indirect band-gap closing is a key signature of the upcoming SDE. Therefore, the AM substantiates the role of an external magnetic field when there exists SOC. {Note that given the indirect nature of the band gap, the same set of parameters for low-energy and
tight-binding models may not give rise to identical effects in both the above models. However, the same physics can be observed after tuning the  parameters suitably in both the low-energy and tight-binding models.}   Having analyzed the band dispersion, we now proceed with the formulation of SDE in the next section.

We note that, in our current implementation, $q$ is treated as a prescribed parameter rather than determined by full self-consistent free-energy minimisation $F(q, \Delta)$. We made this choice primarily to simplify the multi-parameter space while isolating the exact symmetry-breaking role played by finite-momentum pairing. 
Crucially, performing a full self-consistent calculation to find the equilibrium momentum $q_0$, at which  $F(q=q_0, \Delta)$ is minimum, yields the exact same qualitative physics and symmetry-breaking properties as choosing $q$ by hand. The distinction between the two approaches is purely quantitative, while the essential mechanisms, such as the non-zero diode effect, sign structure, and bound-state dynamics, remain qualitatively identical. Note that $q_0$ is fixed and an internally emergent quantity, while $q$ is a variable and an external quantity.

\subsection{Formulation of SDE}
\label{ssec:sde2}

We now discuss the framework to compute the supercurrent. 
There exist two methods, one is using the condensation energy in  momentum space and other one  is based on the current operator formalism in real space. These formalisms apply to both the models with BdG Hamiltonian $\mathcal{H}_{\rm I(II)}$ in momentum  and real spaces. We only follow the condensation energy formalism for our analysis of SDE.  
We first demonstrate the momentum space formalism where we evaluate the derivative of the condensation energy density with respect to the  FF pairing momentum $q$.

The BdG Hamiltonian 
$\mathcal{H}^{\rm I(II)}_k(q,\Delta)$ is diagonalised on a discrete momentum grid, and the free energy density is obtained as \cite{Daido2022, Kinnunen2018, Legg2022}
\begin{equation}
\Omega_{\rm I(II)}(\Delta,q) = -\frac{1}{L}\sum_{k,\eta=\pm} 
\frac{1}{\beta}\ln\!\Big[\cosh\!\left(\tfrac{\beta E^{\rm I(II)}_{\eta,k,q}}{2}\right)\Big],
\label{eq:5}
\end{equation}
where the index $\eta$ labels two branches of the spectrum i.e., valence and conduction bands with  $E^{\rm I(II)}_{\eta,k,q}$ being the positive eigenvalues of $\mathcal{H}^{\rm I(II)}_k(q,\Delta)$.   $\beta=1/T$ is the inverse temperature, and $L$ is the length of the nanowire or the number of $k$ points used for Brillouin-zone sampling $k=2\pi n/L$ with $n=1,\cdots, L$. Hence, the condensation energy density is given by$
F_{\rm I(II)}(\Delta,q) = \Omega_{\rm I(II)}(\Delta,q) - \Omega_{\rm I(II)}(0,q)$.
The supercurrent is  then obtained from the derivative of the free energy \cite{Fu2022} 
\begin{equation}
j_{\rm I(II)}(q) = 2e\,\frac{\partial F_{\rm I(II)}(\Delta,q)}{\partial q},
\label{eq:6}
\end{equation}
where $e$ is the electron charge.
The extrema of the current, 
$j_{\max}$ and $j_{\min}$ yield the 
superconducting diode coefficient, quantifying the degree of non-reciprocity, defined as
\begin{equation}
\gamma_{\rm I(II)} = \frac{j_{\rm I(II)}^{\max}-|j_{\rm I(II)}^{\min}|}{j_{\rm I(II)}^{\max}+|j_{\rm I(II)}^{\min}|}.
\label{eq:7}
\end{equation}
This procedure allows us to determine both the current-momentum relation 
$j_{\rm I(II)}(q)$ and the degree of  nonreciprocal response $\gamma$ with the above prescription.

A current operator formalism is followed to compute the total current using the real-space BdG Hamiltonian $\mathcal{H}^{\rm I(II)}_{\rm BdG}(q)$ in Eq. (\ref{eq:1}). The charge current operator in real space is obtained from the hopping terms in the  tight-binding form  
\begin{equation}
\hat J_{\rm I(II)} = i e \sum_{n,m} 
 \Psi_n^\dagger T^{\rm I(II)}_{nm} \Psi_m  ,
\label{eq:8}
\end{equation}
where $T^{\rm I(II)}_{nm}= (1/2)\big(t [\delta_{n,m+ 1} +\delta_{n+1,m}] \sigma_0 \tau_z - i\lambda (\delta_{n,m+1} -  \delta_{n+1,m}) \sigma_{z(x)} \tau_0 + J_A [ \cos 2\alpha( \delta_{n,m+ 1}+\delta_{n+1,m })  +  i\sin 2 \alpha ( \delta_{n,m+ 1} - \delta_{n+1,m}) ] \sigma_z \tau_z \big) $ 
is the hopping matrix of size $4N\times 4N$ appearing in the BdG Hamiltonian.
The expectation value of the BdG current operator over occupied quasiparticle states yields the supercurrent $j_{\rm I (II)}(q)$,
\begin{align}
j_{\rm I(II)}(q) &= \frac{1}{N}\sum_{E^{\rm I(II)}_n<0} \langle u^{\rm I(II)}_n | {\hat J}_{\rm I(II)} | u^{\rm I(II)}_n \rangle f(E^{\rm I(II)}_n) \notag \\
&= \frac{1}{N}\,\mathrm{Tr}\!\big[ (U^{\rm I(II)}_{\rm occ})^\dagger {\hat J}_{\rm I(II)} U^{\rm I(II)}_{\rm occ} F\big],
\label{eq:9}
\end{align}
where $f(E^{\rm I(II)}_n)=(1+\exp(-\beta E^{\rm I(II)}_n))^{-1}$ is the Fermi-Dirac distribution with Fermi-energy  $E^{\rm I(II)}_F=0$ and $F$ is a diagonal matrix with elements as $f(E^{\rm I(II)}_n)$.  
$|u^{\rm I(II)}_n\rangle$ are eigenvectors of ${ H}^{\rm I(II)}_{\rm BdG}(q)$ Eq.~(\ref{eq:1}), associated with eigenvalue $E^{\rm I(II)}_n$, and $U^{\rm I(II)}_{\rm occ}$ collects the occupied eigenvectors as columns composing a matrix of $4N \times 2N$. The definition of efficiency  Eq. (\ref{eq:7}) is independent of the method adopted to compute supercurrent. 
In the case of tight-binding models,  we note that $j_{\rm I(II)}(q)$, computed using momentum space, yields the same results as obtained from the real space lattice model. In the case of low-energy model, one can only adopt the condensation energy formalism to estimate the supercurrent.


\subsection{Results}
\label{ssec:sde3}

In this section, we present the key outcomes of our numerical analysis on the current and the diode efficiency. We examine two distinct scenarios: the Ising case and the Rashba case namely, Model I and Model II using Eq. (\ref{eq:2}), respectively. For each scenario, we perform calculations using both the low-energy model and the full tight-binding Hamiltonian. A comparative discussion of these results is provided below.

We now examine the variation of the supercurrent $j_{\rm I(II)}(q)$, computed using Eq. (\ref{eq:6}), with FF momentum $q$ for Model I and II in Figs. \ref{fig:5} (a) and \ref{fig:6} (a), respectively. We find that supercurrent passes through $q=0$ for $J_A=0$ in both the models and the momentum $q_*$ at which supercurrent vanishes $j(q_*)=0$ acquires more positive values as $J_A$ increases till a threshold value.  Till this threshold value, $j_{\rm I(II)}(q)$ does not show much asymmetry in terms of its positive and negative values.  
Above this value, for example, $J_A>0.5$ ($1.0$) for Model I (II), $q_*$ suddenly becomes negative and supercurrent shows significant asymmetry  in terms of its positive and negative values. This is maximized for $J_A=1.25$ ($1.43$) in Model I (II), as seen from Figs. \ref{fig:5} (a) and \ref{fig:6} (a) where the positive supercurrent is markedly smaller as compared to the negative counterpart.  Note that supercurrent switches its sign just once till the maximum non-reciprocity arises.  Therefore, there exists an optimum choice of $J_A$ for both the models for which the efficiency is maximized. 
After the above values of $J_A$,  supercurrent displays   multiple crossings over $q$ axis leading to multiple sign reversals of the supercurrent  which can be considered as a signature of AM domination. Therefore, the AM strength can be tuned to maximize the diode efficiency resembling the magnetic field variation. The supercurrent approaches zero when $q$ is large indicating the suppression of superconductivity. 
The current-momentum behavior of both the models is qualitatively similar except for the quantitative value of $J_A$ at which the efficiency is maximum.

We now study the SD efficiency $\gamma$ Eq. (\ref{eq:7}) over $J_A$-$\alpha$ plane for Model I and II in Figs. \ref{fig:5} (b) and \ref{fig:6} (b), respectively. The behavior of $\gamma$ between $\pi <\alpha< 2\pi $ is same as that of between $0 <\alpha< \pi $. The periodicity is $\pi$ as the AM term contains 
$\sin 2\alpha$ and $\cos 2 \alpha$. 
Interestingly, the sign of efficiency can be reversed by tuning $\alpha$ which does not have any analog as far as the external magnetic field is concerned. As anticipated, $\gamma$ varies periodically with $\alpha$, having a period of $\pi$. 
One observation is that the $\gamma=0$ is always true at $\alpha=\pi/4$ and $\alpha=3\pi/4$, and the diode efficiency is anti-symmetric about these two points between $0 \to \ \pi/2$ and $\pi/2 \to \ \pi$ respectively. In Model I, the efficiency acquires a finite value for smaller AM strength as compared to Model II. For $\alpha=n\pi$, $(n+1/2)\pi$ with $n=0,1, \cdots$, the SD efficiency varies from zero to a maximum value within a window of $J_A$ in Model I. This feature is completely absent in Model II where efficiency is always zero irrespective of the choice of $J_A$. Importantly, the sign of $\gamma$ changes around $\alpha=(2n+1)\pi/4$ [$\alpha=n\pi/2$] and with $n=0,1, \cdots$ in Model I [II]. These features are caused by the underlying nature of the SOC whether it is Ising type or Rashba type.  The band of $J_A$ within which $\gamma$ is finite, depends on $\alpha$ in Model I while in Model II the band of $J_A$ is almost insensitive to $\alpha$ except for around $\alpha=n\pi/2$.

We now examine the evolution of $\gamma$ with chemical potential $\mu$ in Figs. \ref{fig:5} (c) and \ref{fig:6} (c), respectively for Model I and II. The efficiency is insignificant as long as $q_*$ is positive in Figs. \ref{fig:5} (a) and \ref{fig:6} (a). As soon as $q_*$ becomes negative, the efficiency acquires finite values at $\mu=0$. Interestingly, the magnitude of $\gamma$ increases as $\mu$ approaches zero from negative side for both the models.  For Model II, the efficiency reduces and vanishes when $\mu$ increases on the positive side. This behavior is not observed in Model I where efficiency can change its sign. Therefore, the nature of the SOC  also dictate the behavior with $\mu$. However, in both the models, efficiency follows a qualitatively similar trend for negative $\mu$ values.

At the end, we here examine the variation of efficiency with SOC strength in Figs. \ref{fig:5} (d) and \ref{fig:6} (d) for Model I and II, respectively. Remarkably, we find that for different values of $\alpha$, $\gamma$ vanishes without Ising SOC in Model I while it remains finite  without Rashba SOC in Model II. This reinforces the fact that non-zero $\alpha$ breaks IS even without SOC. This IS breaking together with TRS breaking due finite $J_A$ cause the SDE to appear. For negative $q_*$ in  Fig. \ref{fig:5} (a), we find that the magnitude of $\gamma$ increases with Ising SOC strength $\lambda$ followed by a decrease in Model I. On the other hand, for Model II, efficiency associated with negative $q_*$ stays almost constant till a certain value of Rashba SOC strength after which its magnitude decreases.   Therefore, there exists an optimum choice of $\lambda$ for both the models for which the efficiency is maximized.

It is important to note that the efficiency reaches a maximum value of $\approx 0.78$ in both the models justifying the use of AM instead of an external magnetic field. 
In our models, the IS is still broken due to SOC even at $\alpha=0$, the AM perfectly replaces the Zeeman field to break the TRS. An important outcome is that, unlike previous observations in the case of Zeeman field with perpendicular spin component \cite{Legg2022}, the two-component AM with parallel spin component is found 
to produce the diode effect. A single component of AM can still produce non-reciprocity when another parallel component arises from the Ising SOC. Instead of having SOCs, we made use of the finite crystal angle $\alpha\ne 0$, yielding two-component AM, to observe the changes in diode efficiency.


\begin{figure}
\includegraphics[width=0.48\textwidth]{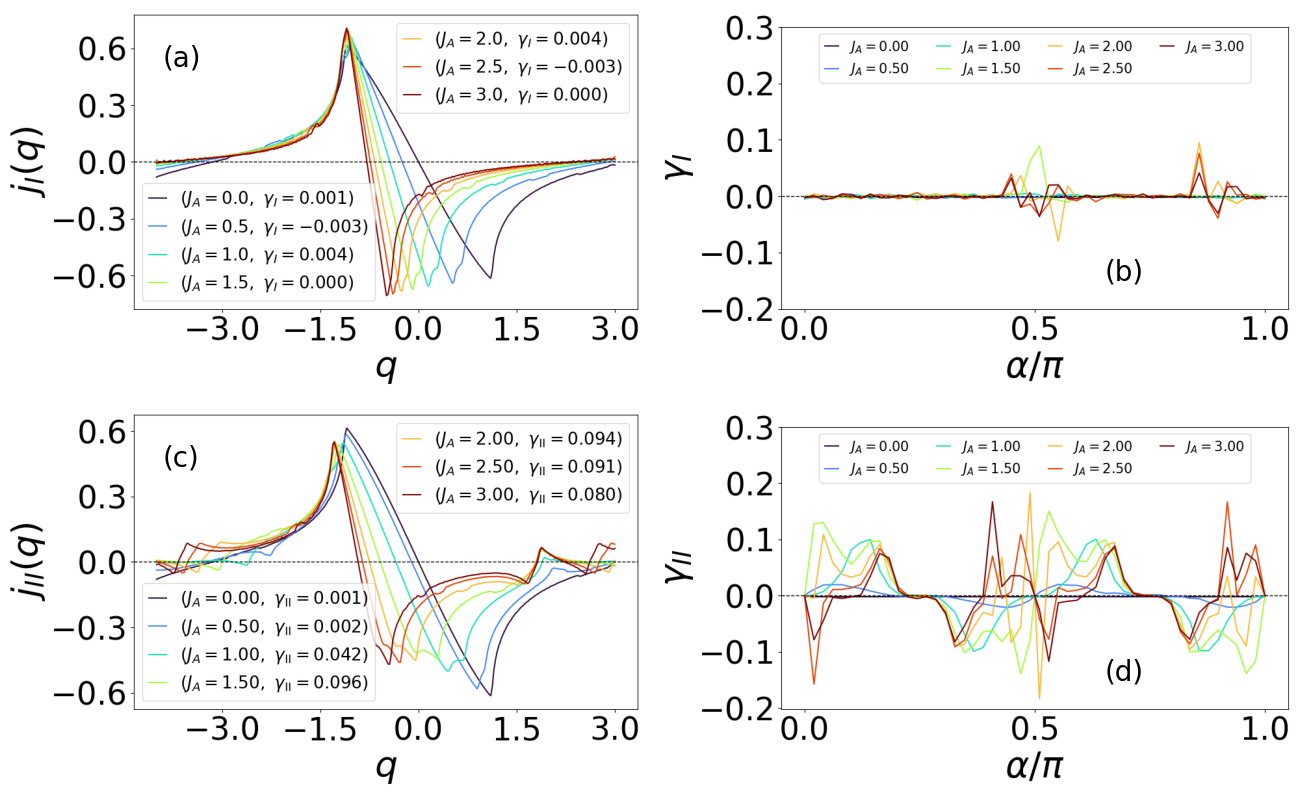}
\caption{We show supercurrent variation with FF pairing momentum in (a) and SD efficiency versus $\alpha$ in (b) for Model I lattice Hamiltonian. We repeat (a,b) for Model II lattice Hamiltonian in (c,d), respectively. We consider $\alpha=0.162\pi$  for (a) and (c).
Parameters: $t=1$, $\mu=0.6$, $\Delta=0.682$, $\lambda=1$, $\beta=10000$. }
\label{fig:7}
\end{figure}


Having investigated the SDE results in low-energy models, we now work with the lattice models to compute the supercurrent following the condensation energy formalism Eq. (\ref{eq:6}) what we did for the low-energy model. We demonstrate the current-momentum behaviour in Fig. \ref{fig:7} (a) and Fig. \ref{fig:7} (c) for Models I and II, respectively. We find that the supercurrent vanishes at $q=0$ when $J_A=0$. Interestingly, in both models, $q_*$ shifts continuously towards more negative values as $J_A$ increases. This shift slows down for higher values of $J_A$. The supercurrent vanishes for $q \to \pm \pi$ indicating the breakdown of superconductivity for larger values of $q$. The variation of $\gamma$ with crystalographic angle $\alpha$ for the lattice model shows qualitatively similar features as that of the low-energy model, i.e., $\gamma$ vanishes at $\alpha=0,\pi/4, \pi/2, 3\pi/4$ and $\pi$. We demonstrate the above variation for Models I and II in  Fig. \ref{fig:7} (b) and Fig. \ref{fig:7} (d), respectively, for better representation. The sign of efficiency can be altered with $\alpha$.
The SD efficiency $\gamma$ is extremely small for Model I, while for Model II it attains a maximum magnitude of $ \approx 0.18$. Model II exhibits noticeable variation of $\gamma$ with $\alpha$ than that of Model I. On the other hand,   Model I and II both exhibit high enough efficiency 
while studying using their low-energy models. In the lattice versions, one can do self-consistent calculations which may lead to almost similar efficiencies. However, the predictions from the low-energy models are qualitatively validated by the tight-binding lattice models. {For instance, from the gap-closing analysis in Figs. \ref{fig:1}, \ref{fig:2},  \ref{fig:3}, and \ref{fig:4} (c,d), when the indirect band gap closes for the inner and outer Fermi rings, the magnitudes of $q_+$ and $q_-$ are significantly different from each other. Noticably, low-energy and tight-binding dispersion can vary quantitatively, resulting in a different set of values of $(q_+,q_-)$ with $q_+\ne q_-$. This possibly refers to the fact that the efficiencies derived from low-energy and tight-binding models are expected to differ, while both of the models suggest the emergence of SDE. }

\begin{figure*}
    \centering
    \includegraphics[width=\textwidth]{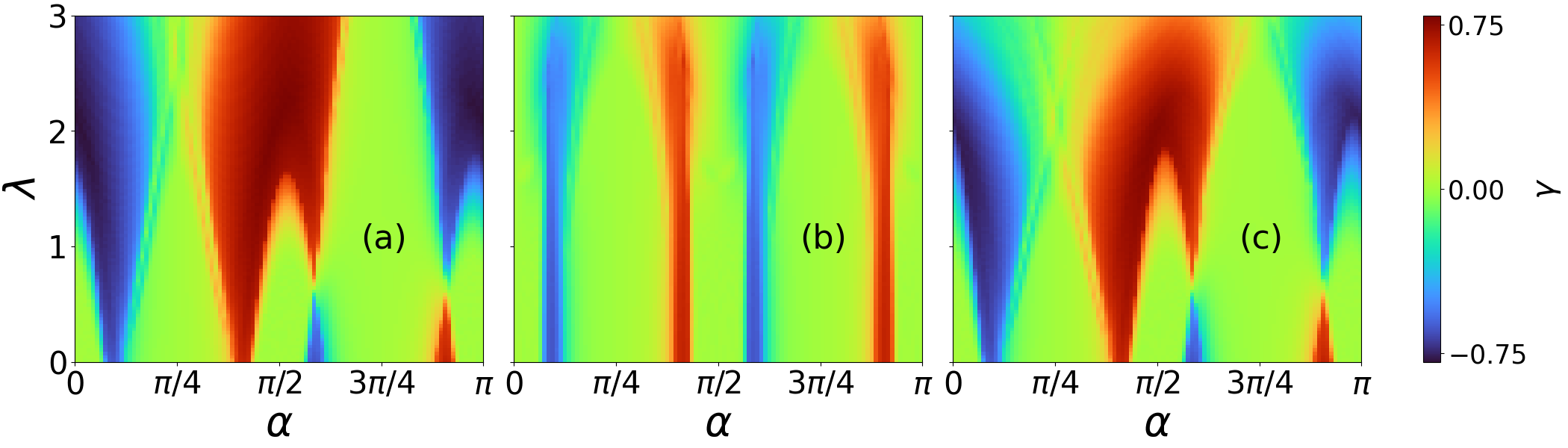}
    \caption{We show the variation of efficiency over the $\alpha$-$\lambda$ plane for low-energy version of Model I in (a). We repeat (a) for Model II in (b). We also repeat the same in (c) with a combination of Ising and Rashba SOC term $\lambda k (\sigma_z+\sigma_x) $ in the Hamiltonian. Parameters: $t=1$, $\mu=-1.52$, $\Delta=2.02$, $J_A=1.22$ and $\beta=10000$. }
    \label{fig:8}
\end{figure*}


We now revert to the efficiency analysis for the  low-energy version of the models, as the efficiency is significantly less for the lattice version of the models. We discuss the behavior of $\gamma$ over $\lambda$-$\alpha$ plane for Model I with Ising SOC and Model II with Rashba SOC, respectively, in Fig. \ref{fig:8} (a) and Fig. \ref{fig:8} (b). 
For $\alpha=0$, one can find SDE in Model I with Ising SOC while SDE no longer exists in Model II with Rashba SOC. There exists a critical strength of Ising SOC above which SDE is experienced for $\alpha=0$.
On the other hand, for $\lambda=0$, SD efficiency acquires finite value for certain window of $\alpha$ around  $\alpha \approx (2n +1) \pi/8$  with $n=0,1,\cdots$. Therefore, crystallographic angle $\alpha$ can break IS and leads to SDE even in the absence of Ising and Rashba SOC. For a given $\alpha  \approx 5 \pi/8 \ne 0$, the sign of efficiency can be altered with Ising SOC while for Rashba SOC such sign reversal is not possible.  For $\lambda=0$, under identical conditions as expected, SDE is obtained for the same windows of $\alpha$ in both the models. {For $\alpha = n \pi /2$ with $n=0,\pm 1, \pm 2, \cdots$, the 1D wire aligns perfectly along high-symmetry crystallographic axes where IS along the wire is preserved, making the $k$-odd $\sin k$ term vanish. On the other hand,  when $\alpha = (2n+1) \pi /4$ with $n=0,\pm 1, \pm 2, \cdots$, the $k$-even $\cos k$  term vanishes, preserving the TRS. Hence, when $\lambda=0$ at these points of $\alpha$, there is no diode effect.}  
Interestingly, the window of $\alpha$ within which efficiency is finite grows with $\lambda$ in Model I justifying  critical strength of Ising SOC. This behavior is not observed in Model II where such a window of $\alpha$ slowly vanishes above a certain value of Rashba SOC. Therefore, Ising SOC and Rashba SOC both produce SDE with different characteristic features.  This proves that AM with the help of the crystallographic angle $\alpha$, can astonishingly replace both the Zeeman field and the SOC to break TRS and IS simultaneously.

We now explore the $\alpha=0$ characteristics of $\gamma$ in more detail. 
For $\alpha=0$, the two-component AM reduces to a single-component AM.
We find that there exists no SDE for $\alpha=0$ in Model II which is markedly different as compared to Model I. 
This is indeed remarkable in the sense that perpendicular spin components i.e., $\sigma_x$ in Rashba SOC and $\sigma_z$ in a single-component AM, do not contribute to SDE irrespective of the strength of Rashba SOC. 
By contrast, the  parallel spin components i.e., $\sigma_z$ in Ising SOC and $\sigma_z$ in a single-component AM, can produce SDE after a threshold value of  Ising SOC. This is a remarkable finding for the 1D field-free SDE while the field-induced SDE requires perpendicular magnetic fields \cite{Legg2022}. Importantly, the complementary momentum functions are needed for the parallel spin components to produce SDE in 1D. This is also evident for $\lambda=0$ case where $\alpha\ne 0$ gives rise to parallel spin component with complementary momentum functions. To strengthen the above inference,  we present a case with the combination of both Ising and Rashba SOC in Fig. \ref{fig:8} (c). In this case, the qualitative features  of $\gamma$ is almost the same as Model I as the parallel spin components are always present. As discussed in Sec. \ref{ssec:sde1}, the parallel spin components break IS and TRS causing the SDE to appear.

In practice, the magnitude of $q$ cannot be arbitrarily large. If $q$ becomes too large, it can strongly perturb the superconducting state and eventually destroy superconductivity. Therefore, in our analysis, $q$ is chosen within a physically reasonable range, generally $q \sim \mathcal{O}(t,\lambda, J_A,\Delta)$, such that superconductivity remains stable while still capturing the effects of finite-momentum pairing. In this sense, $q$ effectively has a natural cutoff determined by the stability of the superconducting phase. Note that after a certain value of $q$, $j(q)$ vanishes as shown in Figs. \ref{fig:5} (a) and \ref{fig:6} (a), indicating the breakdown of superconductivity. Therefore, $q$ should be bounded by the above values.

Given the above discussions, one can anticipate that $q$ is the external probe to account for the finite Cooper pair momentum $q_0$. On the other hand, the SDE can not be examined unless $q$ is applied while SDE can in principle be present even for $q_0=0$ i.e., without any FF pairing. Therefore, external supercurrent alone does not guarantee the emergence of $q_0$ or FF pairing even though it can break IS and TRS. One requires SOC and a magnetic field or altermagnetism to 
stabilize a finite $q_0$ where supercurrent vanishes $j(q_0)=0$ ensuring the emergence of FF pairing.  While MCA stabilizes a finite equilibrium Cooper-pair momentum, $q_0$, leading to an asymmetric $J(q)$ characteristic with $J(q_0)=0$, it is not a necessary condition for the existence of the SDE. Even when a finite momentum $q$ is imposed externally, a finite, albeit much weaker, diode efficiency persists while $J(q=0)=0$. Thus, the SDE is not exclusively tied to the MCA-induced equilibrium momentum $q_0$.

\section{Josephson Diode Effect (JDE)}
\label{sec:jde}

Having studied the AM-mediated SDE, we now focus on the JDE where the Josephson current behaves in an asymmetric manner with the phase difference between the two superconductors. We describe below the AM models to engineer JDE.  

\subsection{Model}
\label{ssec:jde1}

In this section, we analyse three distinct setups to investigate the emergence of the JDE. Specifically, Setup~1 consists of two superconductors, hosting both $s$-wave and $p$-wave gaps, coupled through a quantum wire having SOC and AM. Setup~2 involves two FF superconductors connected via a quantum wire having  SOC and AM. Setup~3 comprises two superconductors and intermediate quantum wire where one superconductor is with FF pairing, SOC and AM, quantum wire hosts only hopping, while the other superconductor only has $s$-wave pairing.

In order to represent the above three  Josephson junction setups concisely, at the beginning we write the lattice Hamiltonian in Nambu/ BdG form.  With the Nambu spinor $\Psi^{\dagger}_n = (c_{n\uparrow}^\dagger,\, c_{n\downarrow}^\dagger,\, c_{n\uparrow},\, c_{n\downarrow})$, the Hamiltonian reads
\begin{equation}
\mathcal{H}_J \;=\; \tfrac{1}{2}\sum_{n,m}\Psi_n^\dagger H^{nm}_J\,\Psi_m,
\quad
H_J =
\begin{pmatrix}
H_N -\mu I  & -i\Delta\sigma_y \\[4pt]
i\Delta^*\sigma_y & -H_N^* +\mu I
\end{pmatrix}.
\label{eq:10}
\end{equation}
The length of the total 1D wire is $L$. The left (right) superconductor has length $L_{l(r)}$. The quantum wire has length $L_{qw}$ such that $L=L_l+L_{qw}+L_r$.
Here the normal such as particle block is decomposed as \(H_N = H_t + H_{\mathrm{SOC}} + H_{\mathrm{AM}}\) where $H_t$ denotes the hopping term.
The $s$ [$p$]-wave pairing block as ${\Delta}_s(n_1,n_2,\phi)=\Delta'_s \exp(i \phi)  \sum_{n,m=n_1}^{n_2} \delta_{n,m}  \sigma_0$ [${\Delta}_p (n_1,n_2,\phi)=i \Delta'_p \exp(i \phi) \sum_{n,m=n_1}^{n_2-1}( \delta_{n,m+1} - \delta_{n+1,m}) \sigma_x$] where $\Delta'_{s[p]}$ indicates the $s$ [$p$]-wave superconducting gap, $\phi$ refers to the phase of the gap; $n_{1(2)}$ denotes the first and last site in the segment of interest. 
In the case of FF superconductor $\Delta(q,n_1,n_2,\phi)= \Delta'_s \exp(i\phi) \sum_{n,m=n_1}^{n_2}e^{iqn} \delta_{n,m} \sigma_0$.  $\mu$ refers to the chemical potential that can be different for superconductors and the quantum wire while $\mu I (n_1,n_2)  = \sum_{n,m=n_1}^{n_2} \delta_{n,m} \sigma_0$. For Ising (Rashba) SOC, $H^{\rm I(II)}_{\mathrm{SOC}}(n_1,n_2) = -\frac{i\lambda}{2}\sum_{n,m=n_1}^{n_2-1}  (\delta_{n,m+1}- \delta_{n+1,m}) \sigma_{z(x)} $. $H_{\rm AM}(n_1,n_2)=J_A \sum_{n,m=n_1}^{n_2-1} [ \cos 2\alpha (\delta_{n,m+1} +\delta_{n+1,m}) +  \sin 2 \alpha (i \delta_{n,m+ 1} - i \delta_{n+1,m}) ] \sigma_z$. 
Note that $H_t$ is present everywhere in the whole 1D wire irrespective of superconductivity, SOC and AM, given by $H_t(1,L)= -t \sum_{n,m=1}^{L-1} (\delta_{n,m+ 1} + \delta_{n+1,m}) \sigma_0$.

The $H_{\mathrm{SOC}}$ and $H_{\mathrm{AM}}$ are present in the intermediate quantum wire (left superconductor) only for setup 1 and 2 (3). The superconductivity $\Delta=\Delta_s \sigma_0 +\Delta_p \sigma_x$  $(\Delta= \Delta_n (q) \sigma_0)$ is present only in setup 1 (2 and 3).  Note that  Ising (Rashba) SOC yields $H^{\rm I(II)}_N$. We adopt a generic representation of Josephson junction   Hamiltonian $ H_{J,x}^y$, following Eq. (\ref{eq:10}), where $y=1,2,3$ denote the setup and $x={\rm I, II}$ stands for the model. {We have demonstrated the schematics of the models and JDE setups in Fig. \ref{fig: Schematic} (b,c), respectively.}
We below demonstrate the Hamiltonian associated with them explicitly. We consider hopping $t$ and  other parameters in the units of meV.

\vspace{6pt}

\textit{\textbf{Setup 1:}} The normal part of the Hamiltonian is given by $H^{\rm I(II)}_N=H_t(1,L)+H^{\rm I(II)}_{\mathrm{SOC}}(L_l+1,L_l+L_{qw}) + H_{\mathrm{AM}}(L_l+1,L_l+L_{qw})$. The chemical potential is given by $\mu I \equiv  \mu_s I (1,L_l) + \mu_0 I (L_l+1, L_l+L_{qw}) + \mu_s I (L_l+L_{qw}+1,L) $. The superconductivity is given by $\Delta={\Delta}_s(1,L_l,\phi) + \Delta_p(1,L_l,\phi) + {\Delta}_s (L_l+L_{qw}+1,L,0)+ {\Delta}_p (L_l+L_{qw}+1,L,0) $. As understood, two different SOC would lead to two different Josephson junction Hamiltonian Eq. (\ref{eq:10}) for the whole composite chain represented by Hamiltonian $H_{J,\rm I(II)}^1$. Note that a similar model has been studied earlier without AM but with external magnetic field \cite{Soori2024}.  The setup 1 would explore the effect of AM on JDE in the presence of various SOCs.

\textit{\textbf{Setup 2:}} In this setup,  the normal part and chemical potential of the Hamiltonian are the same as those of setup 1. The superconductors are in FF state and superconductivity is given by $\Delta=\Delta(q,1,L_l,\phi)+ \Delta(2q,L_l+L_{qw}+1,L,0)$. Likewise setup 1, two different SOC would lead to two different Josephson junction Hamiltonian Eq. (\ref{eq:10})  for the whole composite  chain represented by Hamiltonian $H_{J,\rm I(II)}^2$. The setup 2 allows us to investigate the effect of finite momentum Cooper pair on JDE in addition to the AM and SOCs which are present in the intermediate quantum wire.   Therefore, whether ingredients causing SDE can generate JDE is studied in setup 2.

\textit{\textbf{Setup 3:}} This setup is an admixture of setup 1 and 2. In particular,  the effect of quantum wire is incorporated in the left superconductor which hosts AM and SOC along with FF pairing and there is no $p$-wave superconductivity. The quantum wire does not have SOC and AM in it. It only contains hopping. Meanwhile, the right superconductor only has $s$-wave pairing.  The normal part of the Hamiltonian is given by $H^{\rm I(II)}_N=H_t(1,L)+H^{\rm I(II)}_{\mathrm{SOC}}(1,L_l) + H_{\mathrm{AM}}(1,L_l)$. The chemical potential is given by $\mu I \equiv  \mu_s I (1,L_l) + \mu_0 I (L_l+1,L_l+L_{qw}) +\mu_s I (L_l+L_{qw}+1,L) $. The  superconductivity is given by $\Delta=\Delta(q,1,L_l,\phi)+ \Delta_s(L_l+L_{qw}+1,L,0)$. Likewise setup 1, two different SOC would lead to two different Josephson junction Hamiltonian Eq. (\ref{eq:10})  for the whole composite chain represented by Hamiltonian $H_{J,\rm I(II)}^3$. The setup 3 enables us to examine the role of AM and SOC, being present in the FF superconductor itself, on the JDE without any $p$-wave superconductivity. Therefore, similar to setup 2, we aim here in setup 3 to generate JDE with the help of the ingredients that are essential to cause SDE. To be precise, the left superconductor produces SDE as discussed in the previous Sec.  \ref{sec:sde}. 

\subsection{Formulation of JDE}
\label{ssec:jde2}
In this section, we demonstrate the formalism to compute the Josephson current. 
The total supercurrent through the junction as a function of superconducting phase difference $\phi$ is calculated using the expectation value of the current operator in the many-body ground state of BdG Hamiltonian Eq. (\ref{eq:10}). The current operator $\hat{J}$ is constructed to measure the supercurrent flow, summing over spins, specifically at the bond connecting the superconducting region and the quantum wire/other superconducting region, reflecting the local particle transfer across the interface. It is defined as
\begin{equation}
\hat{J} = i t e \sum_{\sigma}
\left( c_{n+1,\sigma}^\dagger c_{n,\sigma} - c_{n,\sigma}^\dagger c_{n+1,\sigma} \right),
\label{eq:11}
\end{equation}
where $n$ denotes the site at the interface i.e.,  $n=L_l \in$ left superconductor and $n+1 \equiv L_l+1 \in $ quantum wire  for all setups. 
The current operator matrix $\hat{J}$ in the BdG framework is constructed in the Nambu basis to facilitate the particle-hole  degree of freedom. It takes the form
\begin{align}
\hat{J}_{\rm BdG} = 
\begin{pmatrix}
\hat{J} & 0 \\
0 & - \hat{J}^*
\end{pmatrix}.
\label{eq:12}
\end{align}
Note that unlike the case of SDE, the current operator here has a contribution coming only from the  hopping term only which takes care of the interface current. On the other hand, the bulk supercurrent contributes to SDE.

Now coming to the 
BdG Hamiltonian $H_J$ of the complete setups Eq. (\ref{eq:10}), one can obtain the eigenstates $|\psi_n (\phi_l,\phi_r)\rangle$ with eigenvalues $E_n (\phi_l,\phi_r)$ associated with the above Hamiltonian; note that left (right) superconductor has phase $\phi_l=\phi$ ($\phi_r=0$) associated with the gap functions. Therefore, 
$\phi_l-\phi_r=\phi$ denotes the 
net phase difference between left and right superconducting gap functions.    
The total Josephson current with phase difference $\phi$ is then evaluated by
\begin{equation}
J_{\rm total}(\phi) = \frac{1}{et} \sum_{E_n(\phi,0) < 0} \langle \psi_n (\phi,0) | {\hat J}_{\rm BdG} | \psi_n (\phi,0)\rangle f(E_n),
\label{eq:13}
\end{equation}
where the sum is taken over all occupied states below fermi energy $E_F=0$ and $f(E_n)=(1+\exp(-\beta E_n(\phi,0)))^{-1}$ denotes the Fermi-Dirac distribution providing a weight factor associated with each energy level. For ease of notation, considering three setup with two  models,
we designate $J_{\rm total}(\phi)$ as $J^y_{x_{\rm total}}(\phi)$ with $y=1,2,3$ for the setups and $x={\rm I,II}$ for the models. 
To resolve microscopic contributions, the total current is decomposed into two parts corresponding to two  different summations associated with two distinct energy regimes relative to the superconducting gap $\Delta'_s$ as follows
\begin{align}
J_\mathrm{quasi}(\phi) &= \frac{1}{et} \sum_{E_n < -\Delta'_s/2} \langle \psi_n(\phi,0) | \hat{J}_{\rm BdG} | \psi_n (\phi,0)\rangle f(E_n), \notag  \\
J_\mathrm{ABS}(\phi) &= \frac{1}{et} \sum_{-\Delta'_s/2 \leq E_n < 0} \langle \psi_n (\phi,0) | \hat{J}_{\rm BdG} | \psi_n (\phi,0) \rangle f(E_n),
\label{eq:14}
\end{align}
with the total current given by $J_\mathrm{total}(\phi) = J_\mathrm{quasi}(\phi) + J_\mathrm{ABS}(\phi)$. The quasi  [ABS] current $J_\mathrm{quasi}(\phi)$ [$J_\mathrm{ABS}(\phi)$] measures the intraband bulk [interband within the superconducting gap] contributions to Josephson current.

\vspace{0.5em}
The efficiency of the Josephson diode effect is then calculated, which is once again like in Eq. (\ref{eq:7}), quantified by the non-reciprocity parameter ~\cite{Davydova2022,Liu2024} 
\begin{equation}
\gamma^y_x = \frac{J^{y,\mathrm{max}}_{ x_{\rm total}} - |J^{y,\mathrm{min}}_{x_{\rm total}}|}{J^{y,\mathrm{max}}_{x_{\rm total}} + |J^\mathrm{y,min}_{x_{\rm total}}|},
\label{eq:15}
\end{equation}
where $J^{y,\mathrm{max}}_{x,{\rm total}}$ and $J^{y,\mathrm{min}}_{x,{\rm total}}$ denote the maximum forward and reverse total Josephson currents $J^y_{x_{\rm total}}(\phi)$ over the phase period. Here, $y=1,2,3$ denotes three setup and $x={\rm I,II}$ refers to the Model I and II given the setup.  This metric captures the rectification efficiency of the junction. Note that in setups 2 and 3, we choose a particular value of $q$ to analyze the JD efficiency.

Alternatively, the Josephson current,  flowing across the junction in the presence of phase bias $\phi$, can be alternatively computed using the following prescription
\begin{equation}
J_{\rm total}(\phi) = \sum_{E_n(\phi,0) < 0} \frac{\partial E_n(\phi,0)}{\partial \phi} f(E_n)
\label{eq:16}
\end{equation}
The summation over energy levels can be decomposed $E_n < -\Delta'_s/2$ and $-\Delta'_s/2 \le E_n < 0$, to obtain $J_{\rm quasi}$ and $J_{\rm ABS}$, respectively. We follow both the approaches given in Eqs. (\ref{eq:13}), (\ref{eq:16}) to compute the Josephson current and its quasi and ABS contributions. The above correlation persists 
irrespective of the length of the quantum wire and ABS contribution, however, such a correlation breaks once the  $p$-wave superconductivity is present. Note that, we consider $T=0$ such that $f(E_n)=1$ for all our calculations. We follow Eqs. (\ref{eq:13}) and (\ref{eq:14}) to compute the Josephson current for all our following calculations.  

\subsection{Results}
\label{ssec:jde3}


\begin{figure}
\includegraphics[width=0.48\textwidth]{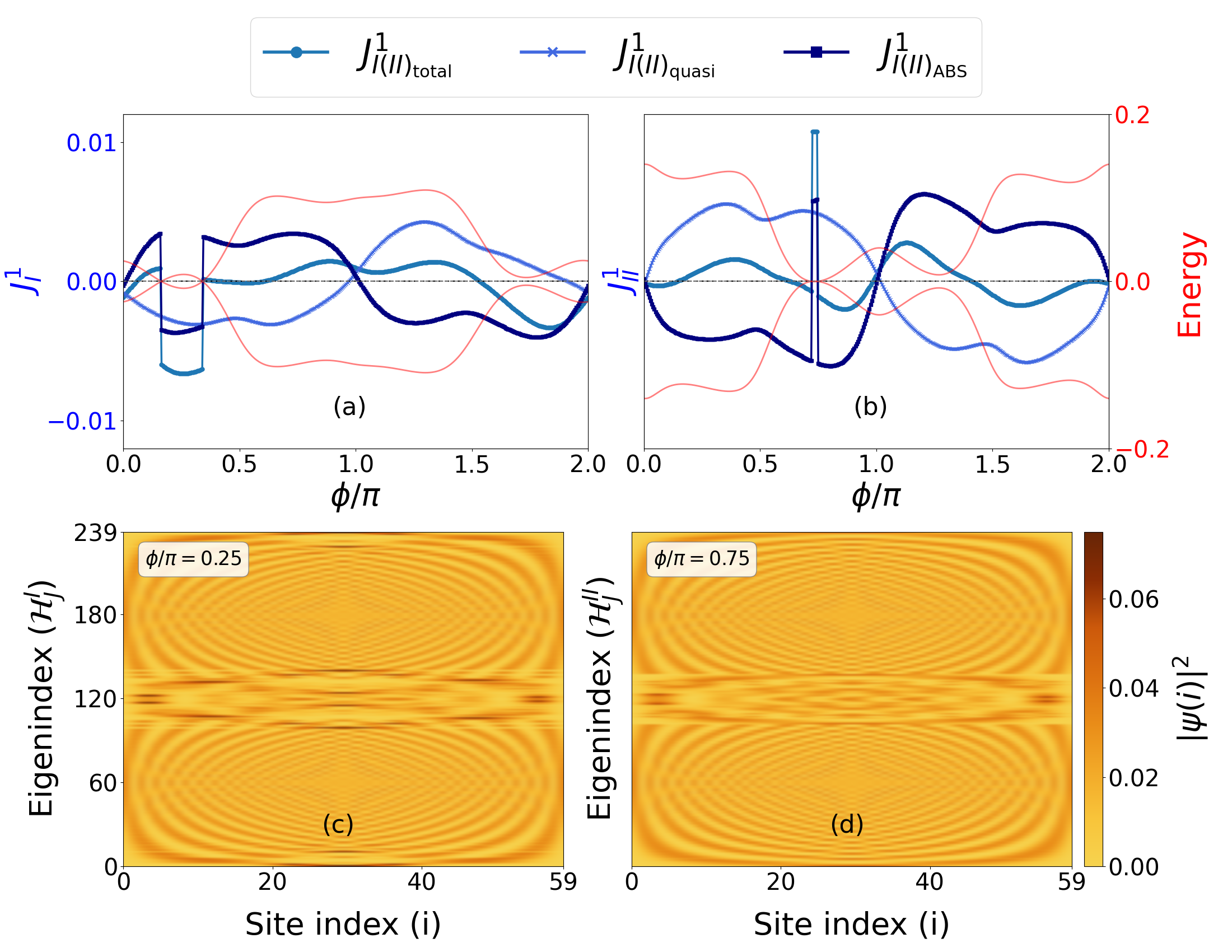}
\caption{We show the variation of supercurrent $J^1_I \equiv J^1_{\rm I_{\rm quasi}}, J^1_{\rm I_{\rm ABS}},J^1_{\rm I_{\rm total}} $ derived from Eqs. (\ref{eq:13}) and (\ref{eq:14}), and energy, with the superconducting phase difference ($\phi$) for setup 1 Model I in (a). We repeat the same for setup 1 Model II in (b). {In (c) and (d), we show the  spatial distributions of the energy eigenstates, obtained after diagonalizing Eq. (\ref{eq:10}) for models I and II in setup 1, respectively.} We consider $J_A = 4.4444$ for (a) and $J_A = 0.6465$ for (b).
    Parameters: 
    $t = 40 ~\text{meV}$, 
    $\mu_s = \mu_0 = -1.875t$, 
    $\Delta_s' = 0.0125t$, 
    $\lambda = 0.04t$, 
    $\Delta_p' = 9.828\Delta_s'$, 
    $\alpha = 0.36\pi$, 
    $L_l = L_r = L_{qw} = 20$.}
\label{fig:9}
\end{figure}



\begin{figure}
\includegraphics[width=0.48\textwidth]{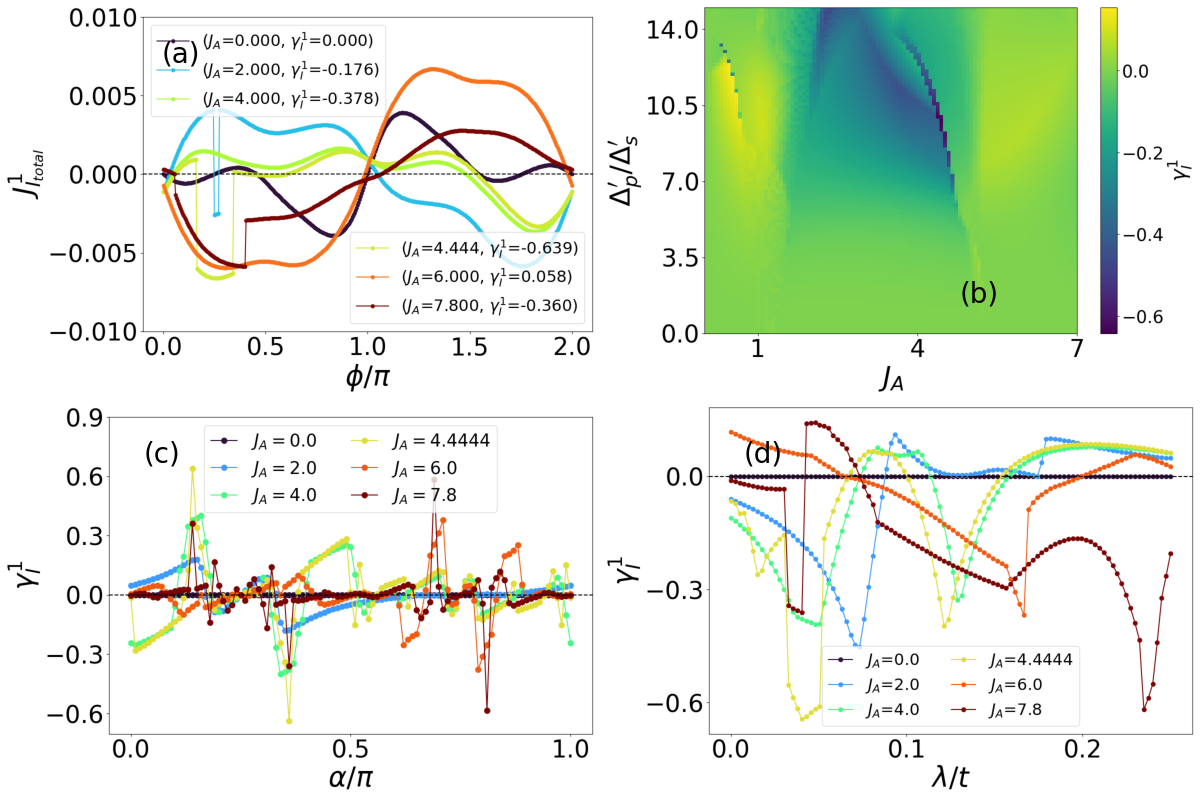}
\caption{In (a), we show the variation of supercurrent $J^1_{\rm I_{total}}$, derived from Eq. (\ref{eq:13}) for setup 1 Model I, with the superconducting phase difference ($\phi$). In (b), we illustrate the heat-map of $\gamma^1_{\rm I}$ derived from Eq. (\ref{eq:15}) as a function of altermagnet amplitude $J_A$, and the ratio of triplet and singlet pairing amplitudes ($\Delta_p'/\Delta_s'$). In (c,d), we display the variation of JD efficiency with $\alpha$ and Ising SOC $\lambda$, respectively. We consider $\lambda = 0.04t$ for (a), (b) and (c). We choose $\Delta_p' = 9.828\Delta_s'$ for (a), (c) and (d). We choose $\alpha = 0.36\pi$ for (a), (b) and (d).
    Parameters: 
    $t = 40\,\text{meV}$,
    $\mu_s = \mu_0 = -1.875t$, 
    $\Delta_s' = 0.0125t$, 
    $L_l = L_r = L_{qw} = 20$.}
\label{fig:10}
\end{figure}




\begin{figure}
\includegraphics[width=0.48\textwidth]{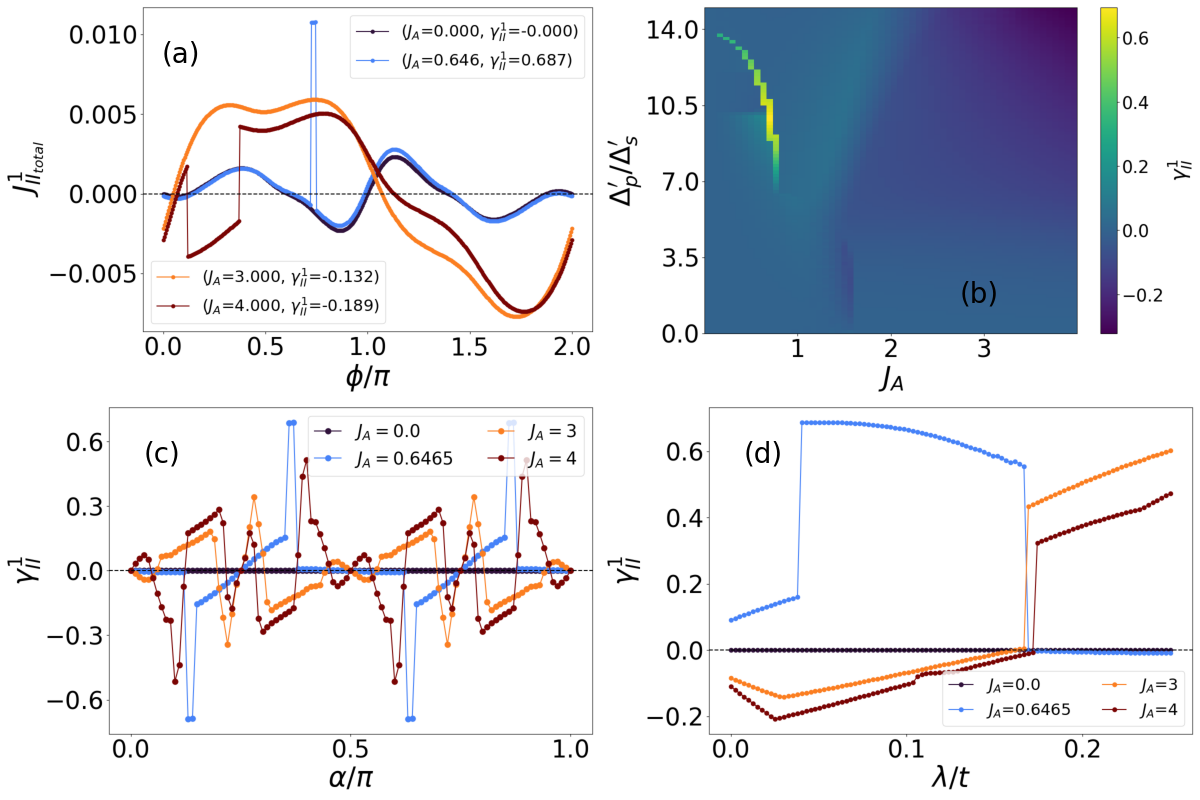}
\caption{In (a), we show the variation of supercurrent $J^1_{\rm II_{total}}$, derived from Eq. (\ref{eq:13}) for setup 1 Model II, with the superconducting phase difference ($\phi$). In (b), we illustrate the heat-map of $\gamma^1_{\rm II}$ derived from Eq. (\ref{eq:15}) as a function of altermagnet amplitude $J_A$, and the ratio of triplet and singlet pairing amplitudes ($\Delta_p'/\Delta_s'$). In (c,d), we display the variation of JD efficiency with $\alpha$ and Rashba SOC $\lambda$, respectively. We consider $\lambda = 0.04t$ for (a), (b) and (c). We choose $\Delta_p' = 9.828\Delta_s'$ for (a), (c) and (d). We choose $\alpha = 0.36\pi$ for (a), (b) and (d).
    Parameters: 
    $t = 40\,\text{meV}$, 
    $\mu_s = \mu_0 = -1.875t$, 
    $\Delta_s' = 0.0125t$, 
    $L_l = L_r = L_{qw} = 20$.}
\label{fig:11}
\end{figure}


In the following sections, we discuss the major outcomes and takeaways from the three setups with underlying Ising SOC and Rashba SOC for Model I and II, respectively in multiple parametric regimes. We examine the total Josephson current into two parts namely quasi- and ABS- contributions to the total current. At the outset, we note that the breaking of IS and TRS 
is important to observe JDE as well. Therefore, MCA is important to produce JDE, however, finite momentum pairing is not thought to be an important ingredient to experience JDE unlike SDE. Below we will explore all setup and examine the role of MCA along with finite momentum pairing. Importantly, the magnetic field is replaced with AM causing the MCA to appear which plays an important role in JDE.

\textbf{\textit{Setup 1:}} We first examine the setup with Model I in Fig. \ref{fig:9} (a) where we show CPR $J^1_{\rm I} \equiv J^1_{\rm I_{\rm quasi}}, J^1_{\rm I_{\rm ABS}},J^1_{\rm I_{\rm total}} $ (left axis) along with the dispersion (right axis) of the  lowest conduction and the highest valence bands. The dispersion inside the superconducting gap 
shows crossing which corresponds to discontinuous profile of 
$J_{\rm ABS}$. This is caused by the change in chirality between two bands during the crossing.  Importantly, the bulk contributions $J_{\rm quasi}$ show smooth continuous behavior due to the gapped nature of the bulk bands. Adding both of them up yields the total Josephson current $J_{\rm total}$ that shows discontinuity when chirality of bands inside the superconducting gap is exchanged during the crossing. Interestingly, there exist asymmetries in the quasi and ABS contributions leading to a substantial non-reciprocity in the CPR for  $J^1_{\rm I_{total}}$ which results in JDE.  These jumps in the current cause unusual critical current values in one of the directions carrying the signature of JDE. 
The bulk contribution can be thought of as a consequence of 
MCA where intraband physics is important  while the ABS contributions are governed by interband physics. Therefore, the 
non-reciprocity in $J^1_{\rm I_{total}}$ is caused by MCA and ABS.

We repeat the above analysis for Model II $J^{1}_{\rm II}  \equiv J^1_{\rm II_{\rm quasi}}, J^1_{\rm II_{\rm ABS}},J^1_{\rm II_{\rm total}} $  in Fig. \ref{fig:9} (b) where  ABS contributions show strong non-reciprocity as compared to quasi contribution. The energy crossings lead to a discontinuity in the ABS contribution indicating the fact that JDE is significantly influenced by ABS rather than MCA. 
We find qualitatively similar behavior of 
$J^{1}_{\rm II}$ as compared to $J^{1}_{\rm I}$. { Comparing Josephson current for Model I and II in setup 1, we find that non-reciprocity in 
$J^1_{\rm I,II_{total}}$ is primarily caused by MCA, while the contributions are more from Ising SOC than from Rashba SOC. }

We now move on to the variation of total Josephson current $J^1_{\rm I_{total}}$ with AM strength $J_A$, crystallographic angle $\alpha$, and SOC strength $\lambda$  for Model I and II in Figs. \ref{fig:10} and \ref{fig:11}, respectively. It can be seen in Fig. \ref{fig:10} (a) that 
there exist no  non-reciprocity 
for $J_A=0$ while  ABS and MCA cause the main non-reciprocity for different non-zero values of $J_A$. Therefore, MCA and ABS both play significant role in producing JDE for Model I. We find that triplet $p$-wave pairing $\Delta_p$ is very crucial to obtain finite JD efficiency $\gamma^1_{\rm I}$ as shown in Fig. \ref{fig:10} (b). The efficiency $\gamma^1_{\rm I}$ is maximized for a certain range of $J_A$ and $\Delta'_p$. The sign of efficiency changes with $J_A$. We investigate the variation of  $\gamma^1_{\rm I}$ with $\alpha$
in Fig. \ref{fig:10} (c) where intermediate (high) values $J_A$ are found to be more relevant to produce JDE around $\alpha=0, \pi/8$, $3\pi/8$, and $\pi/2$ ($5\pi/8$, and $7\pi/8$). Interestingly, $\alpha=\pi/4$, and $3\pi/4$, the efficiency vanishes which resembles the SDE behavior. Importantly, for $\alpha=0$, a single-component AM can generate JDE  in the presence of  Ising SOC and $p$-wave superconductivity. 
We show the variation of efficiency with Ising SOC strength $\lambda$ in Fig. \ref{fig:10} (d) where we find JDE does not exist for $J_A=0$ and $\lambda=0$ suggesting  the importance of MCA. $\gamma^1_{\rm I}$ behaves non-monotonically with $\lambda$. Importantly, intermediate (higher) value of $J_A$ shows maximum efficiency for a relatively smaller (greater) value of $\lambda$. This indicates that there is a relative competition between SOC and AM for JDE to observe.


\begin{figure}
\includegraphics[width=0.48\textwidth]{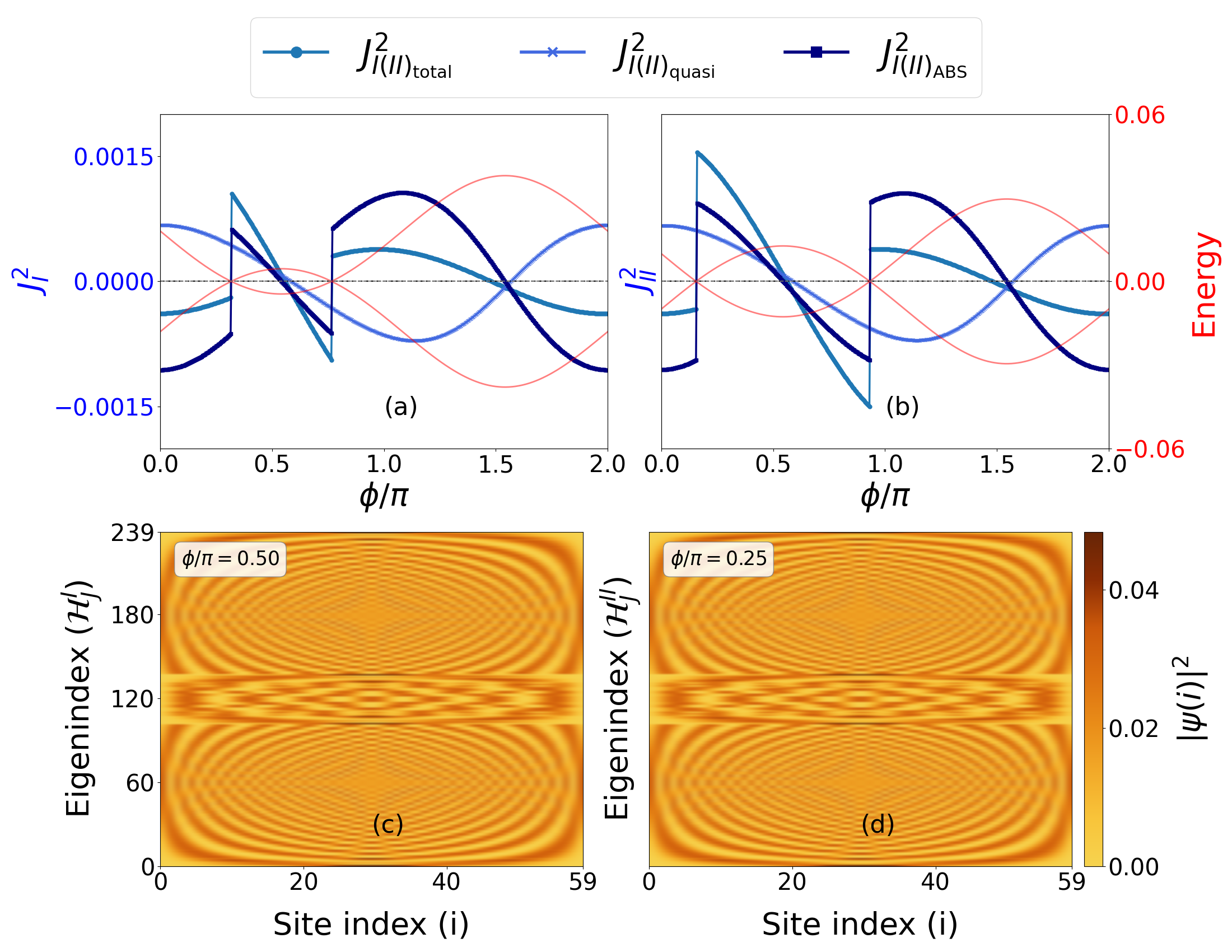}
\caption{We show the variation of supercurrent $J^2_I \equiv J^2_{\rm I_{\rm quasi}}, J^2_{\rm I_{\rm ABS}},J^2_{\rm I_{\rm total}} $ derived from Eqs. (\ref{eq:13}) and (\ref{eq:14}), and energy, with the superconducting phase difference ($\phi$) for setup 2 Model I in (a). We repeat the same for setup 2 Model II in (b). {In (c) and (d), we show the  spatial distributions of the energy eigenstates, obtained after diagonalizing Eq. (\ref{eq:10}) for models I and II in setup 2, respectively.}
    Parameters: 
    $t = 40\,\text{meV}$, 
    $\mu_s = \mu_0 = -1.875t$,  
    $\Delta_s = 0.0125t$, 
    $\lambda = 0.05t$, 
    $J_A=0.5$,
    $\alpha = 0$, 
    $q=0.22$,
    $L_l = L_r = L_{qw} = 20$. }
\label{fig:12}
\end{figure}


\begin{figure}
\includegraphics[width=0.48\textwidth]{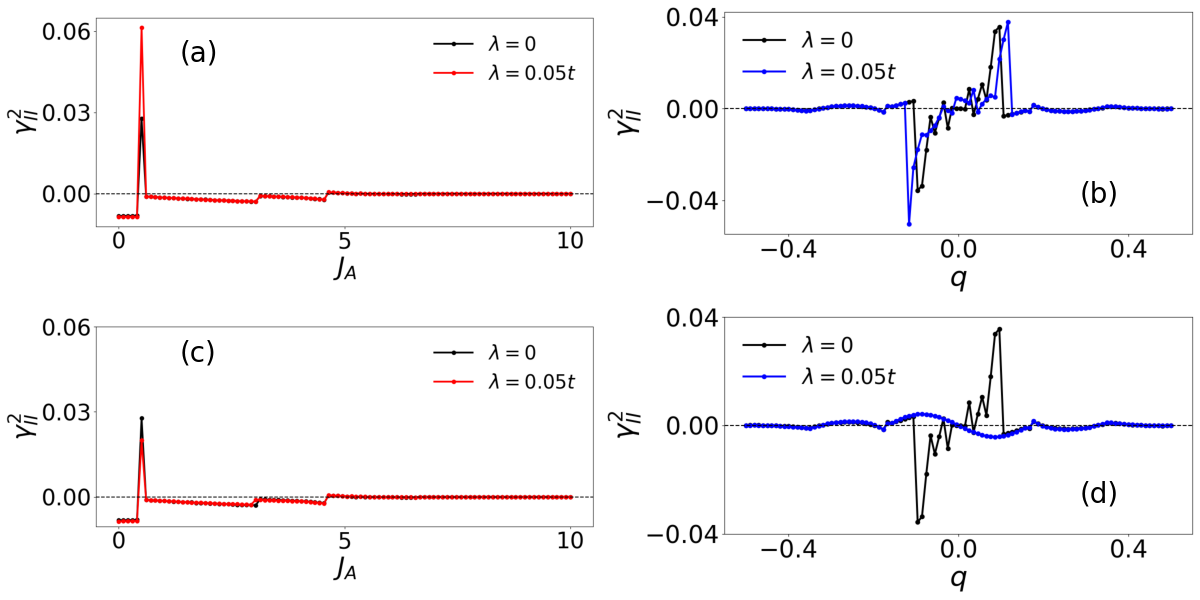}
\caption{In (a,b), we show the variation of JD efficiency $\gamma^2_{\rm I}$, derived from Eq. (\ref{eq:15}) for setup 2 Model I, with altermagnet amplitude $J_A$ and finite momentum of Cooper pair $q$, respectively. We repeat (a,b) for setup 2 Model II in (c,d), respectively. We choose $q=0.22$ for (a) and (c). We consider $J_A=3.13$ for (b) and (d).     Parameters:     $t = 40\,\text{meV}$, 
    $\mu_s = \mu_0 = -1.875t$,
    $\Delta_s = 0.0125t$,
    $\alpha = 0$,
    $L_l = L_r = L_{qw} = 20$. }
\label{fig:13}
\end{figure}


We continue the above analysis for Model II in Fig. \ref{fig:11}.  It is observed from Fig. \ref{fig:11} (a) that the  ABS contribution to diode efficiency is higher than MCA.
From Fig. \ref{fig:11} (b), we can conclude that close to $0.6$ diode efficiency can be achieved for a relatively smaller value of $J_A$  as compared to Model I. Importantly, a smaller region in $\Delta'_p/\Delta'_s$-$J_A$ plane shows significant efficiency which is attributed to the ABS contribution.  This is in contrast to the Model I where ABS and MCA both lead to substantial efficiency in a broader region   over $\Delta'_p/\Delta'_s$-$J_A$ plane. A key outcome from the Fig. \ref{fig:11} (c) is that the $\gamma$ vanishes for values $\alpha=\pi/4,3\pi/4$ while  efficiency peaks around $\alpha=\pi/8, 3\pi/8, 5\pi/8$, and $7\pi/8$ for smaller value of $J_A$. This tendency persists for larger value of $J_A$ as well, however, the efficiency is less. 
Importantly, for $\alpha=0$, a single-component AM can not generate JDE  in the presence of  Rashba SOC and $p$-wave superconductivity. In Fig.  \ref{fig:11} (d), we again show that JDE is absent when $J_A=0=\lambda$. The non-monotonic behavior in $\gamma^1_{\rm II}$ with SOC is a common observation for both the models.

As discussed above in the context of Fig. \ref{fig:10} (c) and Fig. \ref{fig:11} (c) that 
at $\alpha=0$ the diode efficiency is non-zero for Model I, while it is zero for Model II irrespective of the value of $J_A$.  It can also be seen that once $\alpha \neq 0$ in Model II, we have non-zero diode efficiency.
These observations are a direct signature of the JDE being only caused by the parallel spin components i.e., either by two-component AM  or single-component AM in conjunction with Ising SOC, while the $p$-wave superconductivity is present. Note that the  IS and TRS are broken in both of the above instances. 
This suggests that the crystallographic angle $\alpha$ can induce distinct tunability for  the JD efficiency depending upon the  nature of 
SOC.  We note that the JDE was studied earlier with external magnetic field  \cite{Soori2024}. Our work sheds light on the field-free JDE where JD efficiency can be controlled with $J_A$ and $\alpha$ suggesting more tunability in the field-free case.  However, the $p$-wave pairing is very important to obtain JDE even in the field-free case.  Interestingly, the JD efficiency in AM-mediated case is significantly higher than the magnetic field-mediated case \cite{Soori2024}.


\begin{figure}
\includegraphics[width=0.48\textwidth]{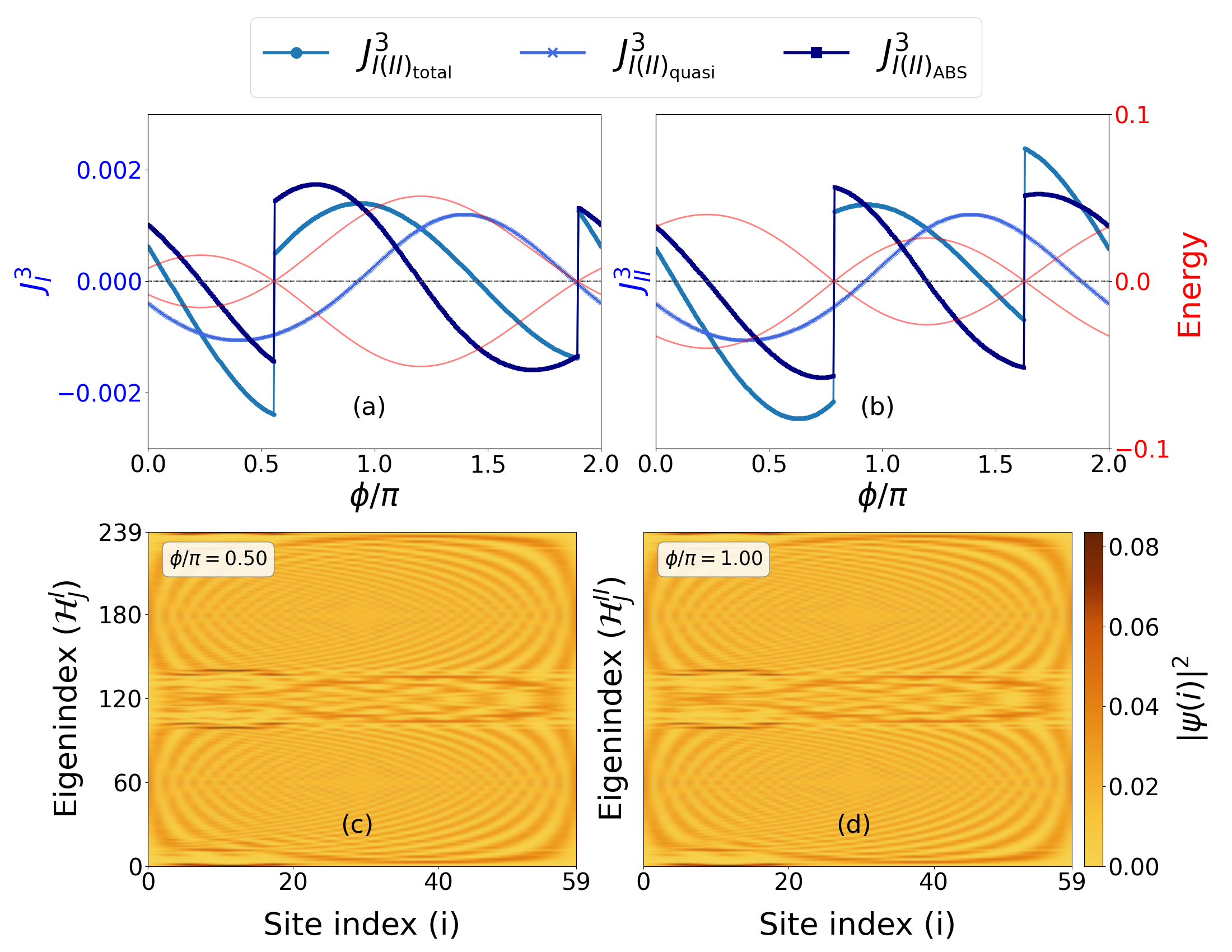}
\caption{We show the variation of supercurrent $J^3_I \equiv J^3_{\rm I_{\rm quasi}}, J^3_{\rm I_{\rm ABS}},J^3_{\rm I_{\rm total}} $ derived from Eqs. (\ref{eq:13}) and (\ref{eq:14}), and energy, with the superconducting phase difference ($\phi$) for setup 3 Model I in (a). We repeat the same for setup 3 Model II in (b). {In (c) and (d), we show the  spatial distributions of the energy eigenstates, obtained after diagonalizing Eq. (\ref{eq:10}) for models I and II in setup 3, respectively.}    Parameters: 
    $t = 40\,\text{meV}$, 
    $\mu_s = \mu_0 = -1.875t$,  
    $\Delta_s = 0.0125t$,
    $\lambda = 0.05t$, $J_A=3.43$, $\alpha=0$, $q=0.34$, 
    $L_l = L_r = L_{qw} = 20$.}
\label{fig:14}
\end{figure}



\begin{figure}
\includegraphics[width=0.48\textwidth]{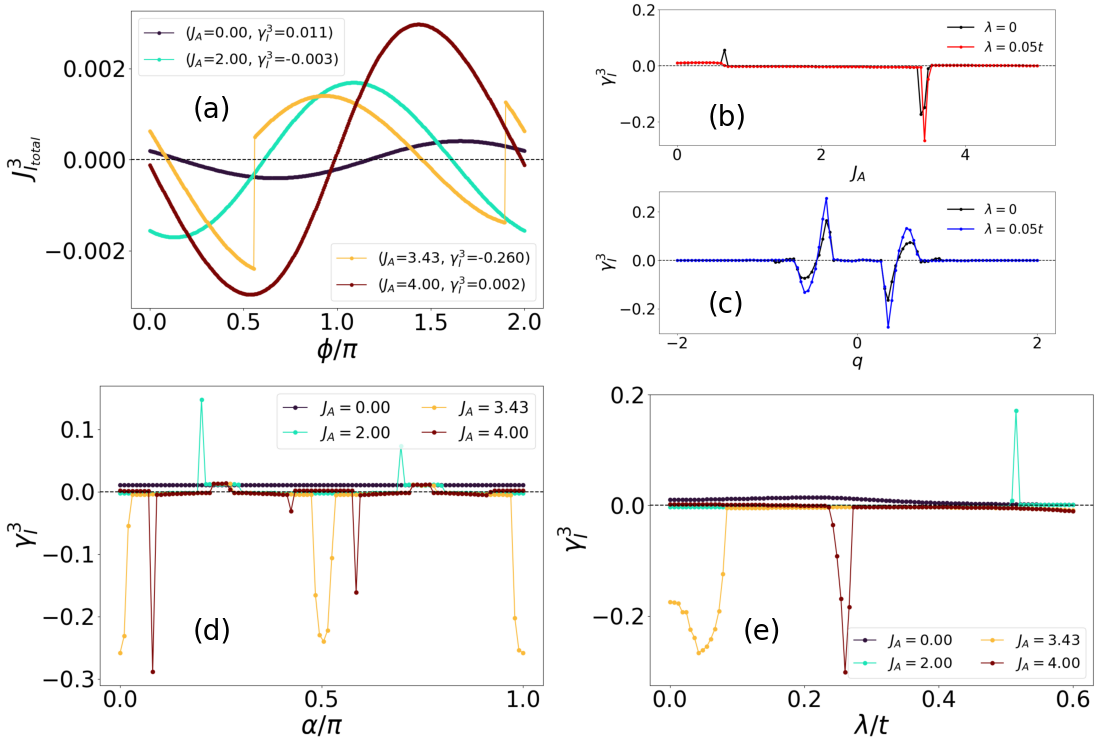}
\caption{In (a), we show the variation of supercurrent $J^3_{\rm I_{total}}$, derived from Eq. (\ref{eq:13}) for setup 3 Model I, with the superconducting phase difference ($\phi$). In (b,c), we illustrate the variation of $\gamma^3_{\rm I}$ derived from Eq. (\ref{eq:15}) as a function of altermagnet amplitude $J_A$, and the finite momentum of Cooper pair $q$, respectively. In (d,e), we display the variation of JD efficiency with $\alpha$ and Ising SOC $\lambda$, respectively.   
We consider  $\lambda = 0.05t$  for (a) and (d). We consider  
    $J_A = 3.43$ for (c). We choose 
    $\alpha = 0$ for (a), (b), (c) and (e). We choose  
    $q=0.34$ for (a), (b), (d) and (e). Parameters: 
    $t = 40\,\text{meV}$, 
    $\mu_s = \mu_0 = -1.875t$, 
     $\Delta_s = 0.0125t$,
    $L_l = L_r = L_{qw}=20$.}
\label{fig:15}
\end{figure}




\begin{figure}
\includegraphics[width=0.48\textwidth]{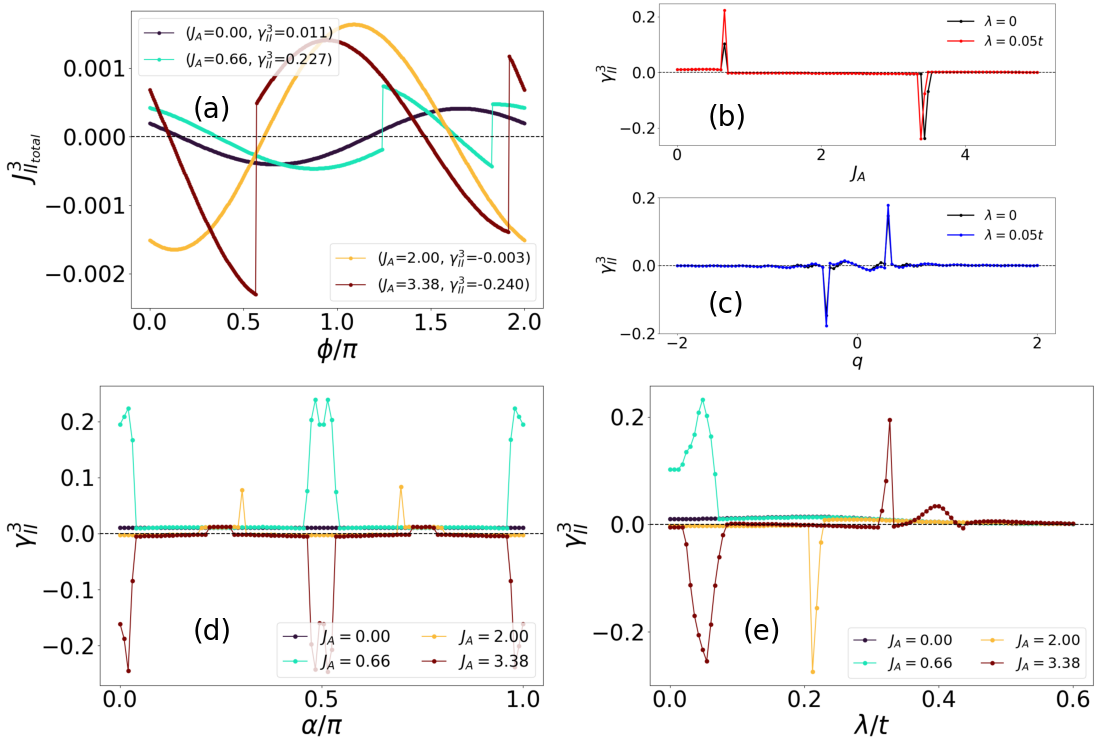}
\caption{In (a), we show the variation of supercurrent $J^3_{\rm II_{total}}$, derived from Eq. (\ref{eq:13}) for setup 3 Model II, with the superconducting phase difference ($\phi$). In (b,c), we illustrate the variation of $\gamma^3_{\rm II}$ derived from Eq. (\ref{eq:15}) as a function of altermagnet amplitude $J_A$, and the finite momentum of Cooper pair $q$, respectively. In (d,e), we display the variation of JD efficiency with $\alpha$ and Rashba SOC $\lambda$, respectively.  
We consider  $\lambda = 0.05t$ for (a) and (d). We consider $J_A = 0.6565$ for (c). We choose      $\alpha = 0.48\pi$ for (a), (b), (c) and (e). We choose      $q=0.34$ for (a), (b), (d) and (e). Parameters: 
    $t = 40\,\text{meV}$, 
    $\mu_s = \mu_0 = -1.875t$, $\Delta_s = 0.0125t$,
    $L_l = L_r = L_{qw}=20$.}
\label{fig:16}
\end{figure}


\textbf{\textit{Setup 2:}} Having demonstrated the setup 1, we now turn our attention to the setup 2
for Model I and Model II in Figs. \ref{fig:12} and \ref{fig:13}, respectively. We find ABS contributions in both the models give rise to JDE as shown in Figs.  \ref{fig:12} (a), and \ref{fig:12} (b), however, finite-momentum does not cause significant non-reciprocity compared to the $p$-wave superconductivity. The efficiency $\gamma^2_{\rm I,II}$ in both models show delta-function like spiky  behavior with $J_A$ see Figs. \ref{fig:13} (a,c). We find an extended region of $q$ within which JD efficiency shows primary peak-dip and secondary oscillations, see 
Figs. \ref{fig:13} (b,d).  Interestingly, efficiency reverses its sign with positive and negative $q$.
Importantly, JD efficiency is finite even for $J_A=0$ and $\lambda=0$ indicating the fact that finite momentum causes IS and TRS breaking that results in MCA. At the same time, this also  
renders the limitations of MCA with $s$-wave superconductivity when AM and SOC are both present in the quantum wire.

\textbf{\textit{Setup 3:}} We first examine the dispersion and the associated Josephson current $J^3_{\rm I,II}  \equiv J^3_{\rm I,II_{\rm quasi}}, J^3_{\rm I,II_{\rm ABS}},J^3_{\rm I,II_{\rm total}} $ in Figs. \ref{fig:14} (a) and \ref{fig:14} (b) for 
Model I and II, respectively. We find that the quasi-contribution $J_{\rm quasi}$ does not show any non-reciprocity while $J_{\rm ABS}$ is the key resource for the non-reciprocity in $J^3_{\rm I,II_{total}}$. The degree of non-reciprocity is more compared to setup 2, advising the fact that the simultaneous presence of SOC, AM with FF superconductivity in the left superconductor suffices for JDE even in the absence of $\Delta_p$ type superconducting gap.  Note that the left superconductor can show SDE as studied in the previous section can now host JDE when it is in contact with the 
quantum wire that only contains hopping. Therefore, MCA along with finite-momentum superconductivity can give rise to JDE under appropriate conditions.

We now turn our attention to the variation of current $J^3_{\rm I,II_{total}}$ and efficiency $\gamma^3_{\rm I(II)}$ with the angle $\alpha$, SOC $\lambda$, AM strength $J_A$ and finite momentum $q$ for Model I and II in Figs.  \ref{fig:15} and \ref{fig:16}, respectively. The ABS contributions to non-reciprocal behavior of current changes as $J_A$ changes  for both models, see Fig.  \ref{fig:15} (a) and Fig \ref{fig:16} (a). For some values of $J_A$, the ABS contribution is completely absent leading to vanishing non-reciprocity. This is markedly different from setup 1 where quasi contribution to Josephson current also shows significant non-reciprocity. Therefore, the 
MCA along with finite momentum  in setup 3 is very effective in generating the non-reciprocity in the ABS contribution. We  find an isolated value of $J_A$ at which $\gamma^3_{\rm I}$ shows a sharp dip, caused by the ABS contribution only, as displayed in Fig. \ref{fig:15} (b). There exists a window  in $q$ for which $\gamma^3_{\rm I}$ remains finite, see Fig. \ref{fig:15} (c). This is similar to case 2, however,  $\gamma^3_{\rm I}$ shows prominent peaks and dips both in either sides of $q=0$ referring to a smooth behavior of efficiency for case 3 as compared to case 2.   Interestingly, the efficiency does not change its sign with $\alpha$ for a given value of $J_A$, see Fig. \ref{fig:15} (d). This is markedly different from setup 1 where $\alpha$ makes the efficiency alter its sign. The magnitude of efficiency increases with $J_A$ as far as the peak or dip values of efficiency are considered.  We find similar results  when efficiency is examined with the strength of Ising SOC $\lambda$, see Fig. \ref{fig:15} (e). Importantly, efficiency is non-zero for $J_A=0=\lambda$ ensuring the presence of MCA for $q\ne 0$. The MCA-mediated non-reciprocity is indeed small while JDE here is mainly composed of  ABS-mediated.

We now perform the same analysis for the Model II in Fig.  \ref{fig:16}. Rashba SOC works similar to  Ising SOC to observe JDE. The non-reciprocity in $J^3_{\rm II_{total}}$ is mainly governed by ABS, see Fig.  \ref{fig:16} (a). We find two isolated values of $J_A$  at which $\gamma^3_{\rm II}$ shows sharp peak and dip, see Fig.  \ref{fig:16} (b).   Efficiency displays a similar pattern with $q$ in  Fig.  \ref{fig:16} (c).  These features are distinct as compared to Model I  as $\gamma^3_{\rm II}$ shows high positive (negative) values for a certain positive (negative)  $q$ value in model I. Efficiency shows high positive and negative (only high negative) values for  certain values of $J_A$ in model II (I). The sign of efficiency does not change with $\alpha$ and $\lambda$ for a given value of $J_A$, see Figs. \ref{fig:16} (d) and (e). There exist dissimilarities with respect to Model I as the maximum value of efficiency does not increase with $J_A$. Therefore, the nature of the SOC plays a significant role in optimizing the response within a given parameter window.

{To better understand the bulk vs. ABS distinction, we have computed the 
spatial distribution of the 
wave-function  \(|\psi_n(i)|^2\), across  the entire systems for all three setups in Figs. \ref{fig:9}, \ref{fig:12}, \ref{fig:14} (c,d), where \(i\) denotes the lattice site index. The $n$-th energy eigenstates $|\psi_n\rangle=(\psi_{\uparrow,e}(i), \psi_{\downarrow,e}(i), \psi_{\uparrow,h}(i), \psi_{\downarrow,h}(i))$ with $i \in \{1,L\}$, are obtained after diagonalizing the setup Hamiltonian Eq. (\ref{eq:10}) for models I and II. 
The total wave-function intensity at site \(i\) is defined as $|\psi(i)|^2=\sum_{\eta}|\psi_{\eta}(i)|^2, \eta\in\{\uparrow,\downarrow,e,h\},$ where \(\eta\) labels the spin and particle-hole components. All three setups exhibit ABSs, as clearly seen in the LDOS profiles. The states corresponding to the ABSs lie near the mid-gap energy values (in this case, the state indices are \(\sim 100\)–\(140\)). Across all three LDOS profiles, these mid-gap states exhibit significantly stronger spatial localization than the other states. Physically, this indicates that the particle and hole components of the ABS wave functions exhibit enhanced spatial localization, reflecting the nature of Andreev reflection at the junction, whereas the remaining states are comparatively more delocalized. }

Upon investigating the three setups, we find that the efficiency is maximized when $p$-wave superconducting gap $\Delta'_p \ne 0$ is substantially high with respect to the $s$-wave superconducting gap $\Delta'_s$. The efficiency $\gamma^1_{\rm I,II}$ is found to be 60\% which is greater than $\gamma^3_{\rm I,II} \sim 20\%$ where finite-momemtum pairing is the key ingredient. Therefore, $p$-wave superconductivity is still found to be more efficient than finite momentum pairing.  
The reason could be attributed to the fact that JDE is caused by MCA and ABS in setup 1 while it is solely caused by ABS in setup 3. Nevertheless, 
setup 3 proves that $p$-wave pairing is not mandatory to observe JDE. ABS has a finite contribution towards non-reciprocity of critical current both in the presence and absence of $p$-wave pairing. 
The JDE requires breaking of IS and TRS thus leading to MCA while ABS is intrinsic in nature. In setup 1, the JDE is absent in the absence of AM and SOC. However, JDE is present in setup 1 when
two parallel spin components are present i.e., either there exists two-component AM  or single-component AM and Ising SOC are simultaneously present. 
This is in complete contrast to the setup 3 where 
JDE persists, with very little efficiency, even in the absence of AM and SOC. Interestingly, finite momentum breaks IS and TRS of the model generating an effective MCA. On the other hand, ABS causes the non-reciprocity when the parity of the  mid-gap states switches. In the setup 3,  the different types of SOC alters the behavior of efficiency with the remaining tuning parameter such as $\alpha$, $J_A$. These changes in the behavior of efficiency are not very evident for setup 1 where $p$-wave superconductivity exists. 
Also, it can be seen that the sign of efficiency can be kept fixed using the FF-state of superconductivity.

Now coming to the comparative study between efficiency, derived from setups 2 and 3, we find that there is a supremacy of efficiency in setup 3 compared to setup 2 i.e., max$\{\gamma^3_{\rm I,II}\}>$ max$\{\gamma^2_{\rm I,II}\}$. The reason  is the following:  The JDE in setup 3 is strongly influenced by the asymmetry between the two superconducting electrodes. The diode effect in setup 3 is caused by the fact that the combination of FF-pairing, AM, and SOC is present in a single superconductor, and the other is just a $s$-wave superconductor. The two superconductors are connected by a normal metal with hopping only. On the other hand, in setup 2  both superconductors are of the FF type, however,  they do not possess AM or SOC which are present only in the connecting quantum wire bridging the two superconductors. As a result, the two superconductors are not very much different from each other in setup 2 while in setup 3, they are substantially different. In order to induce pronounced JDE, 
the superconductor should include AM and SOC as what is noticed for setup 3. For setup 2, superconductors are decoupled from  AM and SOC which is why setup 2 is not able to induce JDE as efficiently as setup 3.   The current transport is more non-reciprocal when MCA ingredients are simultaneously present with FF paring in the superconductor, rather they appear individually which is the case for setup 2.  Hence, the configuration in setup 3 introduces a stronger asymmetry in the pairing structure across the junction, thereby enhancing nonreciprocal transport and increasing the JDE.

The key physical distinction is the nature of the states carrying the supercurrent. In the $p$-wave junction (Setup~1), the Josephson current is primarily governed by the continuum bulk quasiparticle spectrum. Here, the MCA depends on the crystallographic angle of the altermagnet. On the other hand,   the bulk band dispersion is governed by the combined action of $p$-wave pairing and altermagnetism, producing an asymmetric spectrum $E(k)\neq E(-k)$. Since the altermagnetic exchange field explicitly depends on the crystallographic angle $\alpha$, $\alpha$ can cause a change in the sign of the effective MCA as well as the band-bending profile of $E(k)$. As a result,  the evolution of energy bands with the superconducting phase difference clearly hosts the effect of crystallographic angle $\alpha$. 
These could potentially lead to a reversal of the nonreciprocal component of the supercurrent when $\alpha$ varies. To be precise, $E$-$\phi$ behavior changes with angle $\alpha$ in such a way that the $I^{\pm}_c$ alter their signs.  Therefore, the observed sign reversal of the diode efficiency is a direct manifestation of the symmetry of the underlying bulk electronic band structure. The above analogy also applies to the sign change of SDE efficiency with $\alpha$ given the fact that  the condensation energy changes with band energy which strongly depends on  $\alpha$.

In contrast, for the Fulde--Ferrell junctions (Setups~2 and~3), the Josephson current is dominated by Andreev bound states localized at the junction. Their dispersion is primarily determined by the finite Cooper-pair momentum $q$, which introduces an intrinsic phase shift $\phi_0 \propto q$ in the current-phase relation. This phase bias fixes the preferred direction of supercurrent transport independently of the crystallographic orientation. This angle maximally alter  bulk dispersion profile. Note that  here in setups 2 and 3, the bulk non-reciprocity is minimal resulting in an insignificant effect of $\alpha$ on the sign change of efficiency in the above setups.   
Therefore,  varying $\alpha$ modifies the magnitude of the diode response through changes in the spectral asymmetry but does not alter its sign, since the directionality is already locked by the finite pairing momentum.

The numerical results, therefore, reveal two fundamentally different microscopic mechanisms for field-free nonreciprocity: a bulk-band-driven mechanism controlled by the symmetry of the electronic dispersion where altermagnetic angle can change the direction of the current and efficiency sign as observed in JDE setup 1. The same phenomena applies to efficiency associated with SDE as a function of altermagnetic angle irrespective of the choice of the model.  On the other hand, 
an Andreev-bound-state-driven mechanism controlled maximally by finite-momentum pairing and hence the change in the direction of current or efficiency sign is not primarily governed by the altermagnetic angle as seen in JDE setups 2 and 3.
Further, since $p$-wave pairing promotes Majorana bound states (MBSs) that directly reflect the properties of the bulk states, the supercurrent in JDE  Setup 1 is dominated by bulk band contributions rather than purely localised Andreev bound state contributions. {However, we have checked that the electron and hole weights of the states, lying within the superconducting gap,  for a given spin polarization  are not equal. This leads to the fact that there are no MBSs in the system.}

\section{Connection between field-free SDE and JDE}
\label{connection}

In this section, we intend to find the underlying connection between  the SDE and JDE. The MCA causes both the diode effects to appear. As we have shown, two-component AM with finite crystallographic angle  can break both the symmetries and sufficing the role of a magnetic field and SOC. This is why SDE can be observed even in the absence of SOC in both Model I and II.
In other words, the simultaneous presence of  a single-component AM and Ising SOC above a threshold strength could lead to SDE even in the absence of Rashba SOC. Interestingly, the above phenomena are mainly caused due to existence of  parallel spin components with complementary momentum functions. Such an arrangement breaks TRS and IS at the same time. Interestingly, the same arrangement leads to the emergence of JDE in setup 1 in the presence of $p$-wave superconductivity where Rashba SOC is no longer required. 
Similarly, finite momentum of Cooper pair breaks TRS and IS both and results in MCA. This is why JDE is observed in setup 2 and 3 even in the absence of AM and SOC. However, the JD efficiency is significantly suppressed for setup 2  when the superconductor model does not host  FF superconductivity, SOC and AM all simultaneously unlike setup 3 that exhibits noticeable JDE. Therefore, MCA is the regulator of the diode effects while the source of MCA can be external or internal. In our case, all the ingredients are primarily internal and hence SDE or JDE can be obtained while crystallographic angle and chemical potential can result in significant tunability. 
Moreover, the  crystallographic angle plays a crucial role in terms of sign reversal of the efficiency. The variation of this angle can 
be understood as the lattice orientation with respect to its magnetic profile. In short, the parallel spin components with complementary momentum functions along with finite FF pairing are instrumental to experience the SDE in 1D. The 
above ingredients causes JDE to appear as well in 1D even in the absence of  $p$-wave superconductivity.

\section{Conclusion}
\label{conclusion}

\begin{table*}[t]
\centering
\renewcommand{\arraystretch}{1.35}
\begin{tabular}{|p{0.21\linewidth}|p{0.27\linewidth}|p{0.21\linewidth}|p{0.27\linewidth}|}
\hline

\multicolumn{2}{|c|}{\textbf{1D Superconducting Diode Effect (SDE)}} &
\multicolumn{2}{c|}{\textbf{1D Josephson Diode Effect (JDE)}} \\
\hline
\textbf{Existing Literature} &
\textbf{New advancements} &
\textbf{Existing Literature} &
\textbf{New advancements} \\
\hline

Conventional Field-induced 1D SDE relies on the SOC to break IS and on two perpendicular components of the Zeeman field to break TRS \cite{Legg2022}. &
In field-free SDE, AM replaces the two components of the Zeeman Field to generate SDE, along with SOC.&
1D JDE with conventional p-wave superconductors required the SOC and the Zeeman Field in the quantum wire to break IS and TRS, respectively \cite{Soori2024}. &
Field-free JDE can be generated in conventional p-wave superconductors by applying AM and SOC in the quantum wire to break IS and TRS, respectively. \\
\hline

SOC is mandatory to generate SDE by breaking IS. &
AM, along with $\alpha$, can replace both the Zeeman field and the SOC, respectively. &
SOC in the quantum wire was required to generate JDE. &
Two-component AM is enough to produce a finite JDE even in the absence of SOC. \\
\hline

SDE requires perpendicular spin components. &
Parallel spin components with complementary momentum functions generate SDE. &
The microscopic origin of JDE was mainly due to MCA. &
Identifies bulk-driven (in p-wave setup) and ABS-driven (in FF setups) JDE. \\
\hline

\end{tabular}

\caption{Comparison between existing literature and the present work.}
\label{tab:novelty}
\end{table*}

In this work, we have proposed and analysed AM  as a new and effective route to achieve nonreciprocal transport in superconductors. Traditionally, diode effects in superconductors have relied on external Zeeman fields,  breaking TRS in conjunction with SOC that breaks IS. Our results demonstrate that AM can naturally fulfill this role, offering a field-free, tunable, and symmetry-driven pathway to realize the diode effect in a superconductor and in a Josephson junction namely, SDE and JDE, respectively. We consider a crystallographic angle in altermagnetic term such that the diode effects can be studied with the strength of AM $J_A$ as well as angle $\alpha$. To highlight the novelty of the present study, Table~\ref{tab:novelty} summarizes the existing understanding of field-free SDE and JDE alongside the principal advances established in this work.

We start with SDE in 1D where we consider Ising SOC and Rashba SOC to investigate the tunability of SD efficiency with parameters in Model I and II, respectively. The finite-momentum pairing of FF state is a prerequisite to experience SDE while integrating with the MCA. From the band structures of the low-energy BdG versions of the above models, we find that the gap of the inner and outer Fermi surface  closing at different values of probe current is a direct indication of field-free SDE with AM. We find that  the supercurrent changes its sign just once till the maximum non-reciprocity is achieved while varying $J_A$ for both the models. The variation of SD efficiency with chemical potential shows qualitatively similar features.  Interestingly,  
the nature of the SOC nontrivially modifies 
their behavior with $J_A$ and $\alpha$. The sign of efficiency can be reversed by $\alpha$ irrespective of the nature of SOC while there exists SDE at $\alpha=0$ above a threshold value of Ising SOC only. On the other hand, Rashba SOC can not mediate SDE for $\alpha=0$.

Importantly, SDE is found to survive in the absence of any SOC for certain windows of $\alpha$ as long as $J_A \ne 0$. This suggests that parallel spin components with complementary symmetry functions, either arising from a two-component altermagnetic (AM) order or from a combination of single-component AM and Ising spin–orbit coupling (SOC), can generate a superconducting diode effect (SDE) in the presence of finite-momentum pairing, provided that IS and TRS are simultaneously broken. Therefore, the AM alone with the help of finite 
crystallographic angle i.e., two-component AM  can mediate SDE of high efficiency even in the absence of SOC.  We also extend our analysis to tight-binding lattice model where we find qualitatively similar band gap closing, however, the efficiency is found to be small as compared to the low-energy models. {An important distinction between the Zeeman field-mediated SDE and field-free SDE, where AM plays a pivotal, is the following \cite{yuan2022supercurrent}: The AM couples to the electron momentum in a manner similar to SOC, whereas the Zeeman field does not. As a result, altermagnetism generates a stronger asymmetry between left- and right-moving states than the Zeeman-induced case. This enhanced momentum-dependent spin splitting leads to a more pronounced nonreciprocal response. Another key difference lies in the diode's efficiency. In our model, altermagnetism can produce a significantly larger diode efficiency as compared to the field-mediated SDE, and this occurs even for relatively small values of the AM exchange strength $J_A$. }

Having studied the SDE via the altermagnetic route, we continue to explore these models in the context of JDE in 1D, where three distinct setups are considered. AM allows us to investigate field-free JDE in the following setups. 
In setup 1 (2), we have $p$- and $s$-wave (FF) superconductors in contact with an intermediate quantum wire carrying Ising/ Rashba SOC and AM. In setup 3, we consider an FF superconductor with Ising/ Rashba SOC and AM, and an $s$-wave superconductor and both of them are in contact with the intermediate quantum wire containing hopping only. Interestingly, MCA is guaranteed by FF pairing, causing the JDE to appear in setups 2 and 3 irrespective of AM and SOC, while $p$-wave superconductivity in conjunction with AM causes a significant MCA to arise in setup 1, leading to a higher efficiency than that of setups 2 and 3. The quasi-contribution to the Josephson current from bulk bands can be attributed to the MCA while the ABS-contribution is caused by the states lying inside the $s$-wave superconducting gap. We find that JDE in setup 1 contains both quasi- as well as ABS-contributions while for setup 3 it is maximally dominated by ABS-contributions. This also engenders the efficiency in setup 1 to acquire high  value as compared to setup 3. Similar to SDE,  the parallel spin components with complementary momentum functions are also found to be responsible  to secure JDE in the presence of $p$-wave superconductivity. Therefore, setup 1 shows the emergence of JDE in the absence of FF pairing while IS and TRS are broken    by the the parallel spin components. 
Surprisingly, the JD efficiency in setup 2 is negligible as compared to that of setup 3. This is indicative of the fact that the FF superconductor embedding
the properties of the quantum wire namely, AM and SOC is more effective than a FF superconductor without having the above features.  Note that the 
FF superconductor with AM and SOC is able to produce SDE and the same model can lead to JDE when a heterojunction  with another $s$-wave superconductor and quantum wire without AM and SOC  is made to form a Josephson junction.  Importantly, the sign of efficiency changes with the crystallographic angle in setup 1 with $p$-wave superconductivity while such sign reversal is completely absent for setup 3. Therefore, in terms of JD efficiency $p$-wave superconductivity is  more relevant even in the field-free case. On the other hand, FF superconductivity produces ABS-driven JDE with contrasting angle-dependencies.

Overall, our findings highlight altermagnet–superconductor hybrids as ideal candidates for realizing efficient, magnet-free superconducting diodes in 1D. Such systems combine microscopic tunability, material flexibility, and robustness against external perturbations, making them promising for future quantum technologies requiring directional supercurrents. Moreover, the coexistence of SDE and JDE mechanisms uncovered here points toward a unified picture of nonreciprocal superconductivity where finite-momentum pairing, ABS, and band asymmetry act together to produce robust diode behaviour in 1D. 
Looking ahead, our work lays a foundation for future experimental exploration of altermagnetic materials integrated with superconductors. Investigating multi-band effects, thermal influences, non-Hermitian systems, time-driven systems and topological extensions of these systems could further enrich our understanding of non-reciprocity in unconventional superconductors, and perhaps lead to the realization of tunable quantum rectifiers operating in realistic devices. The interplay of Rashba SOC and AM can induce nontrivial topological features in an underlying 
$s$-wave superconductor, thereby opening a route to topological SDE and JDE where the coexistence of Ising SOC, Rashba SOC, and AM in a FF superconductor facilitates the above. 

\textit{Note added}: We note that there exist similar findings on SDE recently, where  Ising, Rashba SOC in the presence of single-component AM in a $s$-wave superconductor can lead to topological superconductivity as well as SDE \cite{pal2025topological}. 

\section{Acknowledgment}
We thank Arijit Saha and Sayak Bhoumik for useful discussions. TN thanks NFSG  “NFSG/HYD/2023/H0911” from BITS Pilani. TN acknowledges the Advanced Research Grant (ARG) from Anusandhan National Research Foundation Grant No. ANRF/ARG/2025/003163/PS.

\end{document}